\documentclass[aps,prl,reprint,superscriptaddress,floatfix]{revtex4-2}
\usepackage{amsmath,amssymb,bm,graphicx}
\usepackage[dvipsnames]{xcolor}
\definecolor{PRLBlue}{RGB}{46,48,146}
\usepackage[
    colorlinks,
    citecolor=PRLBlue,
    linkcolor=PRLBlue,
    urlcolor=PRLBlue,
]{hyperref}
\usepackage{indentfirst}
\graphicspath{{figures/}}

\begin{document}

\title{Entanglement Growth as Transport Across Schmidt Scales}

\author{Shi-Xin Zhang}
\email{shixinzhang@iphy.ac.cn}
\affiliation{Institute of Physics, Chinese Academy of Sciences, Beijing 100190, China}

\author{Shuo Liu}
\email{sl6097@princeton.edu}
\affiliation{Department of Physics, Princeton University, Princeton, New Jersey 08544, USA}

\author{Yu-Qin Chen}
\email{yqchen@gscaep.ac.cn}
\affiliation{Graduate School of China Academy of Engineering Physics, Beijing 100193, China}

\begin{abstract}
Quantum entanglement growth is commonly summarized by a single entropy, obscuring where correlations reside in the exponentially large Schmidt spectrum and how they form. Here, we introduce Schmidt-scale concentration and dominant Schmidt scale, two coordinates that locate the probability maximum across logarithmic windows in ordered Schmidt-rank space. Applied to quenches of a random-field spin chain, these coordinates distinguish rapid transport of the dominant scale to higher Schmidt rank at weak disorder from strongly suppressed transport despite continued logarithmic entropy growth at strong disorder. The disorder-averaged dynamics exhibit an ordered hierarchy: entropy production peaks first, spectral roughness and exact nonlocal magic peak next, and dominant-Schmidt-scale transport becomes typical only after a substantial delay. Moreover, a solvable head--tail model and controlled numerical experiments reveal the physical origin of this hierarchy: the spectral path determines the order of events, local dynamics on active exchange bonds set their early timing, and intra-subsystem many-body dressing further delays dominant-Schmidt-scale transport. These results establish the Schmidt-scale coordinates as powerful dynamical probes for uncovering fine-grained entanglement structures distinguishing entanglement production, entanglement-spectrum reorganization, and dominant-Schmidt-scale transport beyond entropy alone.
\end{abstract}

\maketitle

\textit{Introduction.---} Nonequilibrium entanglement growth records how isolated systems thermalize and local quantum information spreads. Eigenstate thermalization describes the ergodic equilibrium structure~\cite{deutsch1991quantum,srednicki1994chaos,rigol2008thermalization,dalessio2016quantum}, while quantum quenches generate bipartite entanglement ballistically with hydrodynamic and circuit corrections~\cite{calabrese2005evolution,kim2012ballistic,nahum2017quantum,vonKeyserlingk2018operator,rakovszky2018diffusive,zhou2019emergent}. These dynamics can now be directly probed in quantum simulators, including both the growth and saturation of entanglement~\cite{islam2015measuring,kaufman2016quantum}. The von~Neumann entropy $S$ has consequently become the standard measure of entanglement growth.

\begin{figure*}[t!]
\includegraphics[width=0.95\textwidth]{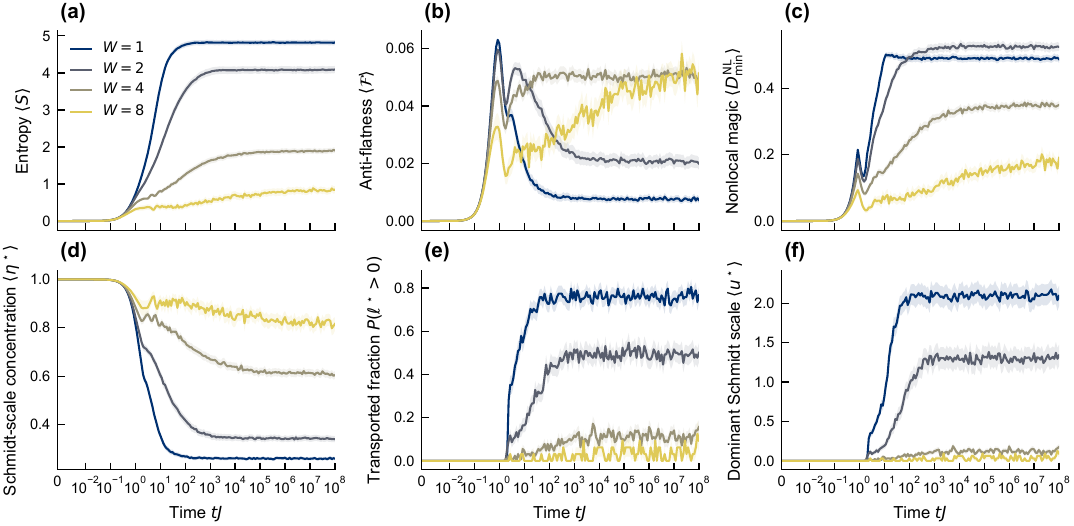}
\caption{\textbf{Schmidt-spectrum dynamics from thermalization to localization.} Exact $L=14$ N\'eel product-state quenches with open boundary conditions at disorder strengths $W=1,2,4,8$ through $tJ=10^8$. Trajectories use 128 disorder realizations for $W=1,2,4$ and 32 for $W=8$; solid curves and shaded bands display disorder means and standard errors. (a) Bipartite entanglement entropy $S$, (b) spectral anti-flatness $\mathcal F$, (c) exact min-relative nonlocal magic $D_{\min}^{\rm NL}$, (d) Schmidt-scale concentration $\eta^\star$, (e) fraction of samples whose dominant Schmidt scale has left the spectral head $P(\ell^\star>0)$, and (f) mean dominant Schmidt scale $\langle u^\star\rangle$. Weak-disorder quenches lose head concentration and transport dominant weight to higher logarithmic rank, whereas strong disorder sustains slow, interaction-driven entropy growth while suppressing that transport.}
\label{fig:dynamics}
\end{figure*}

Yet entropy condenses the exponentially large Schmidt spectrum into one number, obscuring where its weight is concentrated and whether that weight remains local in rank space or is transported between Schmidt scales. The full Schmidt spectrum $\lambda_j$ provides a richer physical landscape. Its level statistics diagnose topological order and criticality~\cite{li2008entanglement,calabrese2008entanglement}; its fluctuations follow induced Wishart ensembles~\cite{lubkin1978entropy,page1993average,marcenko1967distribution,zyczkowski2001induced,sommers2004statistical,nadal2010phase,nadal2011statistical}; and its moments quantify nonclassical computational resources~\cite{liu2022many,sun2025stabilizer,gu2024doped,gu2024zero}. Stabilizer, Pauli, and fermionic diagnostics characterize many-body magic dynamics~\cite{rattacaso2023stabilizer,turkeshi2025pauli,turkeshi2025magic,odavic2025stabilizer,sierant2026fermionic,niroula2024phase,aditya2026mpemba,xiao2026nonstabilizerness}, spectral roughness (anti-flatness) develops a transient barrier during entanglement spreading~\cite{ebner2026magic,zhang2026revealing}, and nonlocal extensions isolate magic in genuine correlations~\cite{qian2025quantum,ding2025evaluating,liu2026nonlocal,collura2026nonlocal}. Schmidt-weight concentration in dyadic windows constrains both exact nonlocal magic and universal entanglement embezzlement~\cite{vanDam2003embezzling,cleve2017perfect,zanoni2024complete,vanLuijk2025critical,sierant2026exact}. However, these properties have usually been studied separately on static states, without a common dynamical description.

In this Letter, we introduce Schmidt-scale concentration and dominant Schmidt scale to construct a unified dynamical description of the entanglement growth. These complementary coordinates locate the most strongly weighted factor-of-two window in ordered Schmidt-rank space. While entropy measures how much entanglement has accumulated, and shape diagnostics record how the spectrum reorganizes, the new quantities reveal where its dominant weight is located and whether it has been transported. Guided by these new observables, we organize the dynamics into one sequence---entanglement production, spectral reorganization, and dominant-Schmidt-scale transport. Thermalization transports the dominant scale to higher rank, whereas many-body localization (MBL) suppresses this transport and keeps the dominant scale near the leading Schmidt weight despite logarithmic entropy growth~\cite{oganesyan2007localization,pal2010many,bardarson2012unbounded,serbyn2013universal,serbyn2013local,huse2014phenomenology,nandkishore2015many,imbrie2016many,schreiber2015observation,zhang2018universal,abanin2019many,falcao2025nonstabilizerness, zhang2026entanglement, liu2025mblmpemba}.

Resolving the onset of transport reveals a robust temporal hierarchy: entropy production peaks first, spectral roughness and nonlocal magic form intermediate maxima, and dominant-Schmidt-scale transport begins only after a substantial delay. To uncover the physical origin, we develop an analytical framework which not only quantitatively predicts the sequence of these events, but also reveals a fundamental dynamical dichotomy: local cross-cut interactions govern the early spectral reorganization, whereas intra-subsystem many-body dressing is responsible for the macroscopic transport. These theoretical insights also explain the dependencies on cut geometry and interactions as well as clarify the protocol-dependent clock relations~\cite{ebner2026magic,zhang2026revealing}.

\textit{Spectral coordinates.---}We investigate the isotropic random-field XXZ spin-1/2 chain,
\begin{equation}
H=J\sum_i\bm S_i\cdot\bm S_{i+1}+\sum_ih_iS_i^z,\quad h_i\sim\mathrm{Uniform}[-W,W].
\label{eq:xxz}
\end{equation}
Here $L$ is the number of sites, $J$ is the exchange scale, and $S_i^\alpha=\sigma_i^\alpha/2$, where $\sigma_i^\alpha$ is a Pauli matrix and $\alpha\in\{x,y,z\}$. We quench N\'eel product states and track the half-chain Schmidt spectrum; time is measured in units of $1/J$. Prior finite-size studies locate the thermal--MBL crossover near $W\simeq3.1$~\cite{zhang2026revealing}. 

For a bipartition of the chain into subsystems $A$ and $B$, with Hilbert spaces $\mathcal H_A$ and $\mathcal H_B$, let $\lambda_j$ denote the ordered eigenvalues of the reduced density matrix $\rho_A$ of subsystem $A$, equivalently the squared Schmidt coefficients. They obey $\lambda_j\geq0$ and $\sum_j\lambda_j=1$ and therefore form a probability distribution, ordered as $\lambda_0\geq\lambda_1\geq\cdots$. The Schmidt rank $r$ is the number of nonzero weights and is bounded by $d_{\min}=\min(\dim\mathcal H_A,\dim\mathcal H_B)$. Because $r$ grows exponentially with subsystem size, we resolve probability on a logarithmic rank axis. Rather than choose only the disjoint shells $1$, $2$--$3$, $4$--$7$, and so on, we evaluate a factor-of-two window at every integer starting rank. The resulting overlapping windows remove dependence on fixed bin edges and resolve the scale more densely; their relation to fixed-shell embezzlement criteria is given in SM Sec.~1. We define the full sliding-octave profile
\begin{align}
\eta_\ell&=\sum_{j=\ell}^{\min(2\ell,r-1)}\lambda_j,\qquad \eta^\star=\max_{0\leq\ell<r}\eta_\ell,\nonumber\\
u^\star&=\log_2(\ell^\star+1),
\label{eq:octave}
\end{align}
with $\ell^\star=\min\{\ell:\eta_\ell=\eta^\star\}$, i.e., the earliest maximizing index. In one-based ranks $q=\ell+1$, the window contains $q,q+1,\ldots,2q-1$ before clipping at Schmidt rank $r$. We call $\eta^\star$ the \emph{Schmidt-scale concentration}, $u^\star$ the \emph{dominant Schmidt scale}, and $(\eta^\star,u^\star)$ the \emph{Schmidt-scale coordinates}. Here $\ell=0$ isolates the leading Schmidt weight ($\eta_0=\lambda_0$), while increasing $\ell$ probes higher logarithmic rank. The coordinate $u^\star$ changes only when a subleading sliding window contains more probability than the leading weight. Representative spectra and a visualization of this logarithmic-rank coarse graining are given in SM Sec.~7~C. %\footnote{See Supplemental Material for observable definitions, model and numerical details, analytic derivations, control protocols, and random-state benchmarks; it includes Refs.~\cite{jasser2026journey,peschel2003calculation,liu2025mblmpemba}}.

We track these coordinates alongside entropy $S=-\sum_j\lambda_j\log_2\lambda_j$, anti-flatness $\mathcal F=P_3-P_2^2$ with moments $P_n=\sum_j\lambda_j^n$~\cite{liu2026entanglementantiflatnessnonlocalnonstabilizerness, ebner2026magic,zhang2026revealing,tnfv-lzfx,karjula2026ebbsflowsquantumlearning}, and min-relative nonlocal magic~\cite{veitch2014resource,howard2017application,leone2022stabilizer,qian2025quantum,sierant2026exact},
\begin{equation}
D_{\min}^{\rm NL}=-\log_2\max_{\substack{k\geq0\\2^k\leq d_{\min}}}\frac{1}{2^k}\left(\sum_{j<2^k}\sqrt{\lambda_j}\right)^2,
\label{eq:magic}
\end{equation}
which isolates genuine nonstabilizerness from local single-qubit magic.
Anti-flatness measures spectral roughness and only vanishes for flat spectrum. Nonlocal magic measures nonstabilizerness: it vanishes if and only if the spectrum is uniform over a dyadic rank $2^k$, and obeys $-\log_2\eta^\star-\log_2(3+2\sqrt2)\leq D_{\min}^{\rm NL}\leq-\log_2\eta^\star$~\cite{sierant2026exact}. Conceptually, these quantities answer three questions: $S$ measures how much entanglement exists, $\mathcal F$ and $D_{\min}^{\rm NL}$ diagnose how its spectrum is shaped, and $(\eta^\star,u^\star)$ locate where its probability is concentrated. We compute the dynamics using exact diagonalization and Chebyshev expansion implemented with TensorCircuit-NG~\cite{talEzer1984accurate,zhang2023tensorcircuit,zhang2026tensorcircuitng}.

\textit{Dynamics across the thermal-MBL crossover.---}Figure~\ref{fig:dynamics} reports the dynamics of the above entanglement-related quantities under the Hamiltonian quench in Eq.~\eqref{eq:xxz}, and it should be read in three groups: panel (a) tracks the amount of entanglement, panels (b) and (c) track entanglement-spectrum reorganization, and panels (d)--(f) track the concentration and transport of dominant weight. At weak disorder ($W=1$), the product state rapidly loses leading Schmidt weight and $u^\star$ moves to higher rank [Fig.~\ref{fig:dynamics}(a), (d)--(f)]. At strong disorder ($W=8$), transport is strongly suppressed: the slow entropy growth is consistent with logarithmic behavior, while the dominant scale remains close to the head over the observed time window. Thus interaction-driven erosion of the leading Schmidt weight can increase entropy over many decades without appreciable dominant-Schmidt-scale transport.

Anti-flatness reveals how the spectral shape changes across the full disorder range [Fig.~\ref{fig:dynamics}(b)]. For every $W$, it initially rises as probability leaving the head forms an uneven tail. At weak disorder it subsequently falls as that weight spreads over many Schmidt modes; increasing disorder prolongs the increasing regime, and the $W=8$ curve shows slow late-time growth without a resolved turnover. 

The scale-sensitive panels separate two processes that entropy alone cannot distinguish. A decrease of $\eta^\star$ while $u^\star=0$ for $t<1$ records erosion of the leading Schmidt weight before any factor-of-two rank window overtakes it. An increase of both $P(\ell^\star>0)$ and $u^\star$ marks the subsequent transport of dominant weight to higher rank. The $W=1$ trajectories display both processes, whereas increasing disorder delays and suppresses transport even as the head continues to lose weight [Fig.~\ref{fig:dynamics}(d)--(f)]. This separation motivates a closer examination of the early-time trajectory.

\begin{figure*}[t]
\centering
\includegraphics[width=0.92\textwidth]{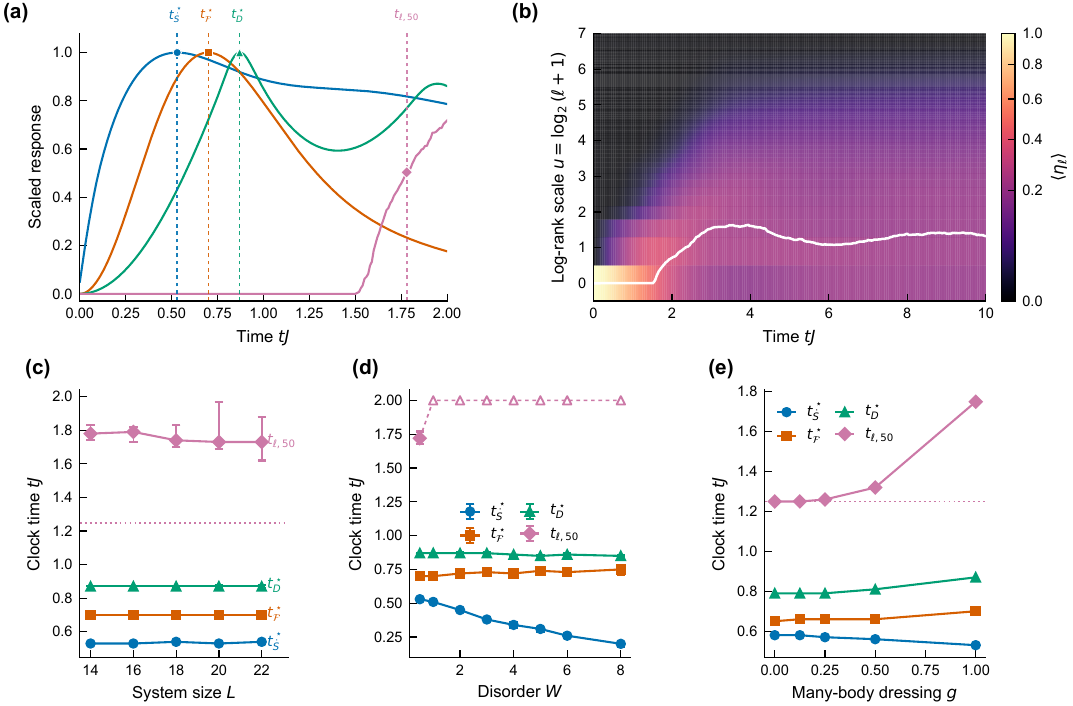}
\caption{\textbf{Four-clock trends.} Disorder-averaged N\'eel product-state quenches with periodic boundary conditions at $W=0.5$, except that (d) varies $W$. (a) Normalized entropy-production rate, anti-flatness, and exact nonlocal magic, plotted with the transported fraction $P(\ell^\star>0)$; dashed lines and markers locate $t_{\dot S}^\star$, $t_{\mathcal F}^\star$, $t_D^\star$, and $t_{\ell,50}$. (b) Complete $L=14$ sliding-window profile through $tJ=10$, with $\langle u^\star\rangle$ in white. Panels (a) and (b) average 256 realizations. (c) The ordering is size stable from $L=14$ to 22. (d) All four clocks versus disorder at $L=14$, using 128 realizations per $W$. The filled diamond is the resolved $t_{\ell,50}$ at $W=0.5$; open triangles and the dashed continuation at the $tJ=2$ boundary denote no majority transport within the simulated time window. In a localized phase, $t_{\ell,50}$ may be infinite. Error bars in (c) and (d) are 95\% trajectory-bootstrap intervals. (e) Clock times when all neighboring bonds within either half are multiplied by $g$ while the two cut bonds remain fixed, using 256 realizations per $g$. The first three clocks vary weakly, whereas dominant-Schmidt-scale transport acquires a pronounced excess delay as the full many-body environment is restored. Pink dotted lines in (c) and (e) mark the isolated-two-bond prediction.}
\label{fig:clocks}
\end{figure*}

\begin{figure}[t]
\centering
\includegraphics[width=0.95\columnwidth]{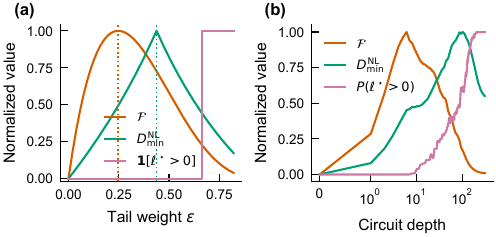}
\caption{\textbf{Spectral thresholds and a circuit control.} (a) Rank-$64$ head-plus-flat-tail spectrum, ordered by increasing tail weight $\epsilon$: normalized anti-flatness $\mathcal F/\mathcal F_{\max}$, normalized exact nonlocal magic $D_{\min}^{\rm NL}/D_{\max}^{\rm NL}$, and the binary transport indicator $P(\ell^\star>0)$. Vertical dotted lines and the indicator jump mark the anti-flatness, magic, and transport thresholds; Eq.~\eqref{eq:headtail} gives their broad-tail limits. (b) Normalized anti-flatness, nonlocal magic, and $P(\ell^\star>0)$, for a periodic $L=12$ brick-wall circuit with weakly entangling $U(1)$-conserving gates, averaged over 128 half-filled product-state realizations.}
\label{fig:mechanism}
\end{figure}

\textit{Four spectral clocks.---}Figure~\ref{fig:clocks}(a) resolves four characteristic times: the time $t_{\dot S}^\star$ at which the entropy-production rate $\dot S$ peaks, the times $t_{\mathcal F}^\star$ and $t_D^\star$ at which $\mathcal F$ and $D_{\min}^{\rm NL}$ reach their first maxima, and the majority-transport time $t_{\ell,50}$, defined as the first time at which at least half the realizations have $\ell^\star>0$. The full octave profile [Fig.~\ref{fig:clocks}(b)] shows the delay: higher-rank octaves acquire weight immediately, but the dominant Schmidt scale remains at $u^\star=0$ until a subleading window outweighs the decreasing head. For quenches with periodic boundary conditions and a N\'eel product-state initial condition, we find
\begin{equation}
t_{\dot S}^\star < t_{\mathcal F}^\star < t_D^\star < t_{\ell,50}.
\label{eq:clock_order}
\end{equation}
These four clocks identify distinct physical features of the same evolving spectrum. We note that the first inequality in Eq.~\eqref{eq:clock_order} does not conflict with the results of Ref.~\cite{zhang2026revealing}, since the two studies use different initial states. The entropy clock $t_{\dot S}^\star$ marks the fastest production of bipartite entanglement. The roughness clock $t_{\mathcal F}^\star$ marks the strongest imbalance between the spectral head and its developing tail, while $t_D^\star$ marks the largest exact nonlocal nonstabilizerness reached along the trajectory. The majority-transport clock $t_{\ell,50}$ is collective: it is reached only when the dominant probability window has left the spectral head in at least half the disorder realizations. Their separation resolves entanglement production, entanglement-spectrum reorganization, and dominant-Schmidt-scale transport as distinct dynamical events.

This clock ordering remains from $L=14$ to $22$ [Fig.~\ref{fig:clocks}(c)] and persists across the disorder scan [Fig.~\ref{fig:clocks}(d)]. Increasing $W$ advances the entropy-rate peak from $t_{\dot S}^\star J=0.53$ to $0.20$, while $t_{\mathcal F}^\star$ and $t_D^\star$ remain near the exchange timescale. Dominant-Schmidt-scale transport is resolved within $t_{\ell,50}J\leq2$ only at $W=0.5$; the upper-edge symbols are finite-window lower bounds. If the asymptotic transported fraction remains below one half, as can occur in the MBL regime, the fourth event is absent and $t_{\ell,50}=+\infty$. These four timescales are features of one evolving Schmidt spectrum: stronger fields render rapid entropy production earlier but leave the two spectral-shape clocks near the local exchange timescale, whereas dominant-Schmidt-scale transport requires a subleading octave window to outweigh the spectral head. The corresponding clock ordering for open-chain protocols is presented in SM Sec.~7~A.

\textit{A unified spectral picture.---}The theory separates a path through spectrum space from the rate at which dynamics traverses it. A minimal head-plus-flat-tail model describes the spectral path: let $\lambda_0=1-\epsilon$ be the head and distribute the remaining weight uniformly among $R$ tail modes, $\lambda_{1\ldots R}=\epsilon/R$. In the broad-tail large $R$ limit,
\begin{align}
S&=h_2(\epsilon)+\epsilon\log_2R,\qquad \mathcal F\to\epsilon(1-\epsilon)^3,\nonumber\\
\eta^\star&\to\max\{1-\epsilon,\epsilon/2\},
\label{eq:headtail}
\end{align}
where $h_2(\epsilon)=-(1-\epsilon)\log_2(1-\epsilon)-\epsilon\log_2\epsilon$ is the binary entropy. Along this broad-tail path, the three shape events occur in the order [Fig.~\ref{fig:mechanism}(a)]
\begin{equation}
\epsilon_{\mathcal F}=1/4 < \epsilon_D=1/2 < \epsilon_\ell=2/3.
\end{equation}
Roughness and magic therefore peak while the head retains substantial probability, whereas dominant-Schmidt-scale transport requires $\epsilon>2/3$ and occurs later. Figure~\ref{fig:mechanism}(a) shows these thresholds in the spectral coordinate $\epsilon$; its purple indicator switches when a tail octave overtakes the head. They become clock times only after a microscopic evolution specifies $\epsilon(t)$ and the evolving tail shape.

The microscopic time dependence is supplied by exchange bonds crossing the bipartition. We call an exchange bond active when it crosses the bipartition, carries a nonzero exchange coupling, and connects antiparallel spins in the initial product state, so that transverse exchange can generate entanglement at first order. The periodic half-chain used in Fig.~\ref{fig:clocks} has two active exchange bonds for the N\'eel state. An independent-two-bond theory model reproduces the ordered roughness and magic clocks and predicts the onset of dominant-Schmidt-scale transport at $t_\ell J=1.231$. Including the random-field detuning across each exchange bond, the same model also predicts the observed advance of $t_{\dot S}^\star$ with increasing $W$: detuning reduces the exchange amplitude but concentrates the resulting entropy production at earlier times. Open-chain results for partitions with one or two active exchange bonds show that equal numbers of active exchange bonds give nearly identical clock times, and random initial states directly sort the early response by the number of active bonds (SM Sec.~7~A).

The independent two-bond model nevertheless predicts dominant-Schmidt-scale transport earlier than the numerical results of the full chain. To explain the discrepancy, we keep the two exchange bonds and random fields fixed while multiplying every XXZ bond within either half by $g$. Thus $g=0$ leaves only the two isolated exchange bonds across the cut and $g=1$ restores the full periodic chain. Figure~\ref{fig:clocks}(e) shows that the transport delay grows continuously as intra-half many-body dressing is restored. Together, these results show that the first three clocks are governed predominantly by local cross-cut dynamics, whereas dominant-Schmidt-scale transport additionally depends on many-body reorganization within the two subsystems (SM Sec.~7~A).

Figure~\ref{fig:mechanism}(b) tests whether the early-time dynamical pattern persists beyond Hamiltonian dynamics. Starting from a half-filled product state, a random circuit with weakly entangling $U(1)$-conserving gates again produces a roughness maximum before dominant-Schmidt-scale transport. A separate SWAP-only circuit exchanges Bell pairs prepared within the two halves and raises $u^\star$ while keeping $\mathcal F=D_{\min}^{\rm NL}=0$, demonstrating dominant-Schmidt-scale transport without roughness or magic. We further examine noninteracting Anderson localization and an interacting l-bit effective model for MBL; their distinct information dynamics are consistent with the theoretical picture developed above (SM Secs.~6 and 7~B).

\textit{Discussion.---}Logarithmic Schmidt rank provides a common transport coordinate for dynamics induced by Hamiltonians and circuits. More broadly, this framework establishes the Schmidt-scale coordinates as highly discriminative dynamical observables, capable of distinguishing distinct nonequilibrium regimes. Furthermore, it captures asymptotic signatures in the thermodynamic limit: whereas $\eta^\star$ converges to a finite constant for Haar-random thermal states, it must rigorously vanish in the context of entanglement embezzling~\cite{vanDam2003embezzling,cleve2017perfect,zanoni2024complete,vanLuijk2025critical} (SM Sec. 8).

As the Schmidt-scale coordinates track both the location and concentration of spectral weight, they may inform adaptive bond-dimension truncation in tensor-network simulations~\cite{schollwock2011density}. Experimentally, tomography and classical-shadow protocols provide routes to estimating entanglement-spectrum information~\cite{huang2020predicting,hu2023classical}, which may enable direct tests of dominant-Schmidt-scale transport.

\begin{acknowledgments}
\textbf{Acknowledgments.---}GPT-5.6 assisted with parts of the code development and analytical derivations. All results were verified by the authors. SXZ was supported by the National Natural Science Foundation of China (No. 12574546), Quantum Science and Technology-National Science and Technology Major Project (No. 2024ZD0301700), and the Chinese Academy of Sciences (No. XDB1680201 and No. YSBR-150).
SL was supported by the Gordon and Betty Moore Foundation through Grant No. GBMF8685 towards the Princeton theory program, the Gordon and Betty Moore Foundation’s EPiQS Initiative (Grant No. GBMF11070), the Global Collaborative
Network Grant at Princeton University, the Simons Investigator Grant No. 404513, the Princeton Global
Network, the NSF-MERSEC (Grant No. MERSEC DMR 2011750), the Simons Collaboration on New Frontiers in Superconductivity (Grant No. SFI-MPS-NFS-00006741-01 and No. SFI-MPS-NFS-00006741-06), the Princeton Catalysis Initiative, the Schmidt Foundation at the Princeton University, European Research Council (ERC) under the European Union’s Horizon 2020 research and innovation program (Grant Agreement No. 101020833). YQC was supported by the National Natural Science Foundation of China (No. 12504599), Quantum Science and Technology-National Science and Technology Major Project (No. 2025ZD0300802), and Science Challenge Project (No. TZ2025017).
\end{acknowledgments}

\bibliography{ref}

\end{document}

\title{Supplemental Material for ``Entanglement Growth as Transport Across Schmidt Scales''}
\author{Shi-Xin Zhang}
\email{shixinzhang@iphy.ac.cn}
\affiliation{Institute of Physics, Chinese Academy of Sciences, Beijing 100190, China}

\author{Shuo Liu}
\email{sl6097@princeton.edu}
\affiliation{Department of Physics, Princeton University, Princeton, New Jersey 08544, USA}

\author{Yu-Qin Chen}
\email{yqchen@gscaep.ac.cn}
\affiliation{Graduate School of China Academy of Engineering Physics, Beijing 100193, China}
\maketitle

\setcounter{secnumdepth}{2}
\renewcommand{\thesection}{\arabic{section}}
\renewcommand{\thesubsection}{\Alph{subsection}}
\renewcommand{\theequation}{S\arabic{equation}}
\renewcommand{\thefigure}{S\arabic{figure}}
\renewcommand{\thetable}{S\arabic{table}}

This Supplemental Material develops the spectral-transport picture used in the Letter. Section~1 defines every observable and clock. Section~2 specifies the Hamiltonian, ensembles, propagation methods, and uncertainty estimates. Sections~3--6 give the analytic structure: general spectral bounds, the head--tail spectrum model, active-bond local dynamics, and interacting l-bit dephasing. Section~7 presents the numerical controls and extended dynamics, and Sec.~8 derives random-state baselines with and without the $U(1)$ constraint. Throughout, Schmidt probabilities are ordered in descending order and logarithms are base two unless $\ln$ is written explicitly.

\section{Spectral coordinates and dynamical clocks}

This section establishes a common notation for the analytic and numerical parts of the work and gives simple reference spectra for each quantity. Consider a normalized pure state across a bipartition $A|B$. In product bases $\{|a\rangle_A\}$ and $\{|b\rangle_B\}$, write $|\psi\rangle=\sum_{ab}C_{ab}|a\rangle_A|b\rangle_B$. A singular-value decomposition (SVD)
\begin{equation}
C=U\,\mathrm{diag}(s_0,s_1,\ldots)\,V^\dagger,
\end{equation}
directly produces the Schmidt amplitudes $s_i$. After sorting them in descending order, the Schmidt probabilities are $\lambda_i=s_i^2$. Equivalently, the $\lambda_i$ are the eigenvalues of $\rho_A=CC^\dagger$ or $\rho_B=C^\dagger C$. Thus every observable used here can be obtained from one SVD of the bipartitioned wavefunction, without choosing a basis within either subsystem. The resulting Schmidt decomposition is
\begin{equation}
|\psi\rangle=\sum_{i=0}^{r-1}\sqrt{\lambda_i}|i_Ai_B\rangle,\qquad \lambda_0\geq\lambda_1\geq\cdots\geq0,\qquad \sum_i\lambda_i=1.
\label{eq:s_schmidt}
\end{equation}
Each $\lambda_i$ is a Schmidt weight: an eigenvalue of either reduced density matrix, equivalently the square of a Schmidt coefficient. Here $r$ is the Schmidt rank, the number of nonzero Schmidt weights, and satisfies $r\leq\min(\dim\mathcal H_A,\dim\mathcal H_B)$. The vectors $|i_A\rangle$ and $|i_B\rangle$ are the left and right singular vectors. The list $\bm\lambda=(\lambda_0,\lambda_1,\ldots)$ is a probability distribution: its overall spread measures entanglement, while its detailed shape records how that entanglement is distributed among Schmidt modes.

Universal entanglement-embezzling families require spectral weight to vanish on every logarithmic rank scale asymptotically~\cite{vanDam2003embezzling,cleve2017perfect,zanoni2024complete,vanLuijk2025critical}. A conventional fixed dyadic partition uses the disjoint one-based rank shells $B_k=\{2^k,\ldots,\min(2^{k+1}-1,r)\}$. We use an origin-independent refinement: for every integer starting rank $q=1,\ldots,r$, the sliding window $I_q=\{q,\ldots,\min(2q-1,r)\}$ spans the same factor of two in rank. Thus neighboring windows overlap, and the fixed shells are precisely the subset $I_{2^k}$. Evaluating every start avoids splitting a concentration peak arbitrarily across fixed bin edges and samples its location more densely on the logarithmic rank axis. In the zero-based indexing of the Schmidt probabilities, the resulting full sliding-octave profile is
\begin{equation}
\eta_\ell=\sum_{i=\ell}^{\min(2\ell,r-1)}\lambda_i,\qquad \eta^\star=\max_{0\leq\ell<r}\eta_\ell,\qquad \ell^\star=\min\operatorname*{arg\,max}_{\ell}\eta_\ell,
\label{eq:s_octave}
\end{equation}
and its logarithmic location is $u^\star=\log_2(\ell^\star+1)$. The window $[\ell,2\ell]$ contains $\ell+1$ consecutive weights until it reaches the end of the spectrum. Its mass $\eta_\ell$ asks how much probability occupies one factor-of-two interval of ordered rank. To relate this refinement to the fixed partition, let $\eta_{\rm part}^\star$ be the largest mass among the disjoint shells $B_k$. Since every $B_k$ is included in the sliding family, $\eta_{\rm part}^\star\leq\eta^\star$. Conversely, any $I_q$ lies within two adjacent fixed shells, so $\eta^\star\leq2\eta_{\rm part}^\star$. Hence $\eta_{\rm part}^\star\to0$ if and only if $\eta^\star\to0$: the two constructions express the same asymptotic embezzlement criterion up to a constant factor, while their finite-rank maxima and maximizing locations need not coincide. We call $\eta^\star$ the \emph{Schmidt-scale concentration}, $u^\star$ the \emph{dominant Schmidt scale}, and $(\eta^\star,u^\star)$ the \emph{Schmidt-scale coordinates}. Large $\eta^\star$ means that one logarithmic rank scale carries much of the state; $\ell^\star$ locates that scale, while $u^\star$ converts multiplicative changes of rank into additive motion. The earliest-maximizer convention removes ambiguity at exact ties.

Three elementary examples demonstrate the convention. A product state has $\bm\lambda=(1)$ and $(\eta^\star,\ell^\star,u^\star)=(1,0,0)$. A Bell pair has $\bm\lambda=(1/2,1/2)$; the two equal octave maxima are resolved in favor of $\ell^\star=0$, so $(\eta^\star,u^\star)=(1/2,0)$. More generally, a flat rank-$2^m$ spectrum has $\eta^\star=1/2$, $\ell^\star=2^{m-1}-1$, and $u^\star=m-1$. A broad flat block can therefore sit at large $u^\star$.

We compare the Schmidt-scale coordinates with three complementary spectral functions. We write $S\equiv S_A=S(\rho_A)$ for the bipartite von Neumann entropy of subsystem $A$. The von Neumann and R\'enyi entropies are
\begin{equation}
S=-\sum_i\lambda_i\log_2\lambda_i,\qquad S_n=\frac{1}{1-n}\log_2P_n,\qquad P_n=\sum_i\lambda_i^n.
\label{eq:s_entropies}
\end{equation}
The von Neumann entropy $S$ measures the total effective number of populated Schmidt modes: $S=0$ for a product state and $S=m$ bits for a flat rank-$2^m$ spectrum. R\'enyi entropies tune the sensitivity to large probabilities; in particular, $S_2=-\log_2P_2$ is determined by the purity $P_2$. Entropy can grow either because probability leaves the largest Schmidt value or because an already existing tail spreads over more modes, a distinction used in Sec.~4.

Anti-flatness is the moment combination~\cite{ebner2026magic,zhang2026revealing}
\begin{equation}
\mathcal F=P_3-P_2^2.
\label{eq:s_af}
\end{equation}
It is the variance of the random variable that takes value $\lambda_i$ with probability $\lambda_i$, and therefore measures eigenvalue inhomogeneity with strong weight near the Schmidt head. It vanishes if and only if the spectrum is flat on its nonzero support, allowing any number of zero-probability tail entries, and is positive otherwise. Anti-flatness therefore resolves a transient roughness barrier that entropy alone cannot locate.

Nonlocal nonstabilizerness isolates the magic that cannot be removed by local unitaries and, for pure states, is governed by the entanglement spectrum~\cite{qian2025quantum}. Let $d_{\min}=\min(\dim\mathcal H_A,\dim\mathcal H_B)$. For each allowed dyadic-rank exponent $k=0,1,\ldots,\lfloor\log_2d_{\min}\rfloor$, define the fidelity to a flat rank-$2^k$ Schmidt spectrum as $F_k$. The exact min-relative version and its optimizing exponent $k_D^\star$ follow directly from the ordered Schmidt amplitudes~\cite{sierant2026exact},
\begin{equation}
F_k=\frac{1}{2^k}\left(\sum_{i=0}^{2^k-1}\sqrt{\lambda_i}\right)^2,\qquad
k_D^\star=\min\operatorname*{arg\,max}_{0\leq k\leq\lfloor\log_2d_{\min}\rfloor}F_k,\qquad
F_{\rm NL}=F_{k_D^\star},\qquad D_{\min}^{\rm NL}=-\log_2F_{\rm NL}.
\label{eq:s_magic}
\end{equation}
The optimization compares the state with flat Schmidt spectra of ranks $1,2,4,\ldots$ allowed by the smaller Hilbert-space dimension; $2^{k_D^\star}$ is the rank of the closest such spectrum, and the smallest exponent resolves an exact tie. We set $\lambda_i=0$ for $i\geq r$. Consequently, $D_{\min}^{\rm NL}=0$ for a product state, a Bell pair, or any stabilizer state. It can remain nonzero even when anti-flatness vanishes: a uniform rank-three spectrum embedded in dimension at least four has $S=\log_2 3$, $\mathcal F=0$, and $D_{\min}^{\rm NL}=\log_2(4/3)$. This example makes the distinction precise. Anti-flatness detects unequal Schmidt probabilities, whereas nonlocal magic detects departure from the entire family of dyadic-flat spectra. Zero padding to the full bipartite dimension leaves $\eta^\star$, $\ell^\star$, and $F_{\rm NL}$ unchanged.

The five observables therefore read the same entanglement spectrum in complementary ways. Entropy measures total spread; anti-flatness emphasizes head--tail unevenness; nonlocal magic measures the distance from the best dyadic-flat sector; the Schmidt-scale concentration $\eta^\star$ measures how concentrated the probability remains within one logarithmic rank interval; and the dominant Schmidt scale $u^\star$ records which interval dominates. For a head-plus-tail spectrum, entropy can already be large while $u^\star=0$ because the leading probability still outweighs every subleading window. This pinned-but-growing regime is one of the central dynamical distinctions resolved in the Letter.

For a disorder or circuit ensemble, angular brackets denote the arithmetic mean over independent realizations. We define four clocks from ensemble-averaged curves: $t_{\dot S}^\star$, $t_{\mathcal F}^\star$, and $t_D^\star$ are the first interior local maxima of the smoothed curves $d\langle S\rangle/dt$, $\langle\mathcal F\rangle$, and $\langle D_{\min}^{\rm NL}\rangle$, respectively, while $t_{\ell,p}$ is the first sampled grid point at which $P(\ell^\star>0)\geq p$. A boundary maximum is unresolved. The Letter uses $p=1/2$ and writes $t_{\ell,50}$. These definitions distinguish a response peak from a threshold-crossing event.

\section{Models, time evolution, and statistical analysis}

This section specifies the physical protocols, propagation algorithms, spectrum extraction, and statistical postprocessing used in every figure. The random-field XXZ model is
\begin{equation}
H=\sum_i\left(S_i^xS_{i+1}^x+S_i^yS_{i+1}^y+S_i^zS_{i+1}^z+h_iS_i^z\right),\qquad h_i\sim\mathrm{Uniform}[-W,W],
\label{eq:s_xxz}
\end{equation}
with $S_i^\alpha=\sigma_i^\alpha/2$ and exchange scale $J=1$. The fields are drawn independently for every disorder realization and then held fixed throughout the evolution. Unless a protocol is named explicitly, the initial state is the half-filled N\'eel product state and the bipartition divides the chain into equal halves. A periodic-boundary-condition (PBC) half chain has two bonds crossing the bipartition, whereas an open-boundary-condition (OBC) half chain has one. The matched-cut protocol instead chooses a central interval inside an open chain, giving two entanglement boundaries without changing the global boundary condition. Random-product protocols sample computational-basis configurations uniformly subject to the same fixed total magnetization; they are initial-state ensembles within the XXZ model.

The disorder regimes can be matched directly to the thermal-MBL finite-size crossover reported for the same random-field XXZ chain in Ref.~\cite{zhang2026revealing}. That work uses Pauli matrices,
\begin{equation}
H_\sigma=\sum_i(\sigma_i^x\sigma_{i+1}^x+\sigma_i^y\sigma_{i+1}^y+\sigma_i^z\sigma_{i+1}^z)+\sum_i h_i^{(\sigma)}\sigma_i^z,
\end{equation}
and finds a crossover near $W_\sigma\simeq6.2$ for its studied sizes. Since $S^\alpha=\sigma^\alpha/2$, the two conventions obey $H_\sigma(W_\sigma=2W)=4H(W)$ for matched disorder samples. They therefore have identical eigenstates at $W_\sigma=2W$, while their times satisfy $t_\sigma=t/4$. The quoted crossover maps to $W\simeq3.1$ in Eq.~\eqref{eq:s_xxz}. We accordingly use $W=1$ as a clear thermal reference, regard $W=2$ as thermal-leaning, place $W=3$--$4$ in the finite-size crossover region, and use $W=8$ as a strong-localization reference. Boundary-condition and finite-size differences make these regime labels more appropriate than a new precision estimate of $W_c$.

For long-time XXZ data we use complete exact diagonalization in the fixed-total-magnetization sector. If $H|E_n\rangle=E_n|E_n\rangle$ and $c_n=\langle E_n|\psi_0\rangle$, every requested state is reconstructed as
\begin{equation}
|\psi(t)\rangle=\sum_n c_ne^{-iE_nt}|E_n\rangle.
\end{equation}
The eigensystem is formed once for each disorder sample, and arbitrary observation times are obtained by changing the phases in this spectral representation.

The dense early-time data through $L=22$ use a Chebyshev expansion of $e^{-iHt}$~\cite{talEzer1984accurate}. Rigorous bounds $E_{\min}$ and $E_{\max}$ rescale the Hamiltonian to $H'=(H-b)/a$ with $a=(E_{\max}-E_{\min})/2$ and $b=(E_{\max}+E_{\min})/2$. We then evaluate
\begin{equation}
e^{-iHt}|\psi_0\rangle=e^{-ibt}\left[J_0(at)|\psi_0\rangle+2\sum_{n=1}^{M-1}(-i)^nJ_n(at)T_n(H')|\psi_0\rangle\right],
\end{equation}
where $T_{n+1}(H')|\psi_0\rangle=2H'T_n(H')|\psi_0\rangle-T_{n-1}(H')|\psi_0\rangle$. We choose $M\geq\lceil at_{\max}\rceil+40$ and reuse the same Chebyshev vectors for all requested times. For the periodic N\'eel data through $tJ=2$ in Fig.~2 of the Letter, $M=64$--$80$ over $L=14$--22 and the maximum saved-state norm deviation is $3.6\times10^{-15}$; the separate $L=14$ extension through $tJ=10$ agrees with fixed-sector exact diagonalization to a maximum absolute octave-profile error of $1.14\times10^{-11}$. The Hamiltonian action is assembled from TensorCircuit-NG Pauli operators~\cite{zhang2023tensorcircuit,zhang2026tensorcircuitng}.

At each saved time, the state amplitudes are reshaped into the coefficient matrix $C_{ab}$ defined above. Singular values are computed separately in the allowed subsystem-charge blocks and then merged and sorted. This blockwise SVD is algebraically identical to an SVD of the full coefficient matrix because total charge makes $C$ block diagonal after a basis permutation; it reduces memory without discarding any Schmidt value. Equations~\eqref{eq:s_octave}--\eqref{eq:s_magic} are evaluated realization by realization, and only then averaged. This order preserves nonlinear quantities such as $\eta^\star$ and $D_{\min}^{\rm NL}$; evaluating them from an averaged spectrum would define a different observable.

The early Hamiltonian grid is uniform with $\Delta t=0.01$. For each observable we first form its ensemble-mean curve and apply a third-order Savitzky--Golay polynomial over 21 consecutive points. The same local polynomial gives $d\langle S\rangle/dt$ analytically for the entropy-rate clock. We define $t_{\dot S}^\star$, $t_{\mathcal F}^\star$, and $t_D^\star$ as the first interior local maxima of the smoothed curves; a boundary maximum is reported as unresolved rather than interpreted as a peak. The transported fraction is computed directly as $P(\ell^\star>0,t)=N^{-1}\sum_s\mathbf 1[\ell_s^\star(t)>0]$, without smoothing, and the transport time $t_{\ell,p}$ is its first grid point at or above $p$. For brick-wall circuits, even--odd layers create staircase plateaus, so the barrier time is defined by the global maximum of the smoothed depth profile rather than by the first local plateau.

Uncertainty estimates resample complete realizations, preserving all correlations among times and observables. Each bootstrap replica draws $N$ trajectories with replacement, recomputes the ensemble curves, repeats the smoothing and peak or threshold extraction, and contributes one clock value; the reported intervals are the 2.5 and 97.5 percentiles of 1000 replicas. Shaded bands on long-time trajectories are standard errors of the realization mean.

The small-angle $U(1)$ circuit is built from independent number-conserving two-site gates
\begin{equation}
U_{ij}=e^{-i\theta\zeta_0}\oplus e^{-i\theta\bm n\cdot\bm\sigma}\oplus e^{-i\theta\zeta_2}
\quad\text{in}\quad
\{|00\rangle\}\oplus\{|01\rangle,|10\rangle\}\oplus\{|11\rangle\},
\label{eq:s_circuit_gate}
\end{equation}
where $\bm n$ is sampled uniformly on the unit sphere and $\zeta_0,\zeta_2$ are independent standard normal variables. Fresh gates are applied on all even bonds and then all odd bonds, including the closing bond, to form a periodic brick wall. The chain begins as one-particle dimers contained entirely within the two halves; a randomly selected fraction $f=0,1/2,$ or $1$ of these dimers is prepared as $(|01\rangle+|10\rangle)/\sqrt2$, while the rest remain in $|01\rangle$. Thus $f$ changes the pre-existing intra-half entanglement reservoir without placing entanglement across the measured cut. We set $\theta=0.25$, so each near-identity gate has weak entangling power, and independently resample every gate. The SWAP relocation control starts from the full-dimer state and applies a SWAP across each of the two bipartition boundaries.

The noninteracting localization control is the open random-field XX chain
\begin{equation}
H_{\rm A}=\sum_{i=1}^{L-1}\left(S_i^xS_{i+1}^x+S_i^yS_{i+1}^y\right)+\sum_{i=1}^{L}h_iS_i^z,
\qquad h_i\sim\mathrm{Uniform}[-W,W].
\label{eq:s_anderson}
\end{equation}
After the Jordan--Wigner transformation its one-particle hopping is $1/2$ and its onsite potential is $h_i$ up to an irrelevant constant. We reconstruct the complete many-body Schmidt spectrum from the subsystem correlation eigenvalues~\cite{peschel2003calculation}. For the interacting localization control, we use $J_{ij}=J_0s_{ij}c_{ij}e^{-|i-j|/\xi}$, where the signs $s_{ij}=\pm1$ are equiprobable and the order-one prefactors $c_{ij}$ are independent samples from $\mathrm{Uniform}[0.5,1.5]$. The numerical Hamiltonian omits the one-body terms $h_i\tau_i^z$ because they factor into unitaries acting wholly within $A$ or $B$ and therefore leave the Schmidt spectrum invariant. These controls compare product-state spectrum formation, relocation of a prepared flat spectrum, localization without dephasing, and interaction-induced dephasing.

\section{General geometry of Schmidt-scale concentration}

The entanglement-related observables obey useful bounds before any dynamical assumption is imposed. These bounds define the allowed region for the numerical trajectories and clarify how concentration constrains nonlocal magic.

Let $M=\lceil\log_2(r+1)\rceil$ and write the fixed shells introduced in Sec.~1 in zero-based form as $B_j=\{2^j-1,\ldots,\min(2^{j+1}-2,r-1)\}$ for $j=0,\ldots,M-1$. These sets are disjoint, their union is $\{0,\ldots,r-1\}$, and each $B_j$ is exactly the sliding window beginning at $\ell=2^j-1$, possibly clipped at the upper end. If $m_j=\sum_{i\in B_j}\lambda_i$, then $\sum_jm_j=1$. The pigeonhole principle therefore gives $\max_jm_j\geq1/M$, and maximization over all sliding windows can only increase the mass:
\begin{equation}
\eta^\star=\max_\ell\eta_\ell\geq\max_jm_j\geq\frac1M.
\end{equation}
The remaining terms follow from a separate chain of elementary inequalities. The $\ell=0$ window contains only the largest Schmidt probability, so $\eta^\star\geq\eta_0=\lambda_0$. Because every $\lambda_i\leq\lambda_0$ and $\sum_i\lambda_i=1$,
\begin{equation}
P_2=\sum_i\lambda_i^2\leq\lambda_0\sum_i\lambda_i=\lambda_0.
\end{equation}
Finally, monotonicity of R\'enyi entropies gives $S\geq S_2=-\log_2P_2$, or equivalently $P_2\geq2^{-S}$. Combining the dyadic-partition argument with $\eta^\star\geq\lambda_0\geq P_2\geq2^{-S}$ yields
\begin{equation}
\eta^\star\geq\max\left\{\frac1M,\lambda_0,P_2,2^{-S}\right\}.
\label{eq:s_eta_bounds}
\end{equation}
For example, when $r=8$, the four blocks are $\{0\}$, $\{1,2\}$, $\{3,4,5,6\}$, and $\{7\}$, so some octave must contain at least one quarter of the total probability. Ordering prevents exact equality of all block masses at finite rank, but the $1/M$ floor is approached asymptotically by slowly varying spectra such as $\lambda_i\propto1/(i+1)$; the purity and entropy steps are saturated by spectra that are flat on their nonzero support. Equation~\eqref{eq:s_eta_bounds} thus combines a rank-geometric floor with familiar head, purity, and entropic floors.
The octave theorem for exact nonlocal magic~\cite{sierant2026exact} bounds the optimal dyadic-prefix fidelity directly in terms of the most massive octave:
\begin{equation}
\eta^\star\leq F_{\rm NL}\leq(3+2\sqrt2)\eta^\star.
\end{equation}
Applying the decreasing function $-\log_2$ reverses the inequalities and gives the first two parts below. Equation~\eqref{eq:s_eta_bounds} supplies the remaining upper bounds: $\eta^\star\geq1/M$ implies $-\log_2\eta^\star\leq\log_2M$, $\eta^\star\geq P_2$ implies $-\log_2\eta^\star\leq S_2$, and $\eta^\star\geq2^{-S}$ implies $-\log_2\eta^\star\leq S$. Combining the steps yields
\begin{equation}
-\log_2\eta^\star-\log_2(3+2\sqrt2)\leq D_{\min}^{\rm NL}\leq-\log_2\eta^\star\leq\min\{\log_2M,S_2,S\}.
\label{eq:s_magic_bounds}
\end{equation}
The maximum possible nonlocal magic therefore grows at most as $\log_2\log_2r$ even for spectra whose entropy grows as $\log_2r$.

Ordering also constrains the position of a concentrated octave. Fix $\ell\geq1$ and abbreviate its mass by $m=\eta_\ell$. Since the $\ell+1$ entries inside the window satisfy $\lambda_i\leq\lambda_\ell$, their sum obeys $m\leq(\ell+1)\lambda_\ell$, or $\lambda_\ell\geq m/(\ell+1)$. The $\ell$ preceding entries $\lambda_0,\ldots,\lambda_{\ell-1}$ are each at least $\lambda_\ell$, so their total mass is at least $\ell m/(\ell+1)$. Normalization then gives
\begin{equation}
1\geq \sum_{i=0}^{\ell-1}\lambda_i+\eta_\ell
\geq \frac{\ell}{\ell+1}m+m
=\frac{2\ell+1}{\ell+1}m.
\end{equation}
Solving for $m=\eta_\ell$ yields
\begin{equation}
\eta_\ell\leq\frac{\ell+1}{2\ell+1}.
\label{eq:s_location_bound}
\end{equation}
This bound approaches $1/2$ as the window moves deeper into the spectrum: a far-tail octave cannot carry substantially more than half of all probability because ordering forces a comparable amount of weight to lie before it. Apply Eq.~\eqref{eq:s_location_bound} to the maximizing window $\ell=\ell^\star$. When $\eta^\star>1/2$, multiplying by the positive denominator and collecting the terms proportional to $\ell^\star$ gives
\begin{equation}
\eta^\star(2\ell^\star+1)\leq\ell^\star+1
\quad\Longrightarrow\quad
(2\eta^\star-1)\ell^\star\leq1-\eta^\star.
\end{equation}
Division by $2\eta^\star-1>0$ therefore gives
\begin{equation}
\ell^\star\leq\frac{1-\eta^\star}{2\eta^\star-1}.
\label{eq:s_location_inverse}
\end{equation}
Because $\ell^\star$ is an integer, the right-hand side may be replaced by its floor. For example, $\eta^\star>2/3$ forces $\ell^\star=0$, while $\eta^\star>3/5$ permits at most $\ell^\star=1$. Strong concentration therefore confines the dominant window quantitatively to the Schmidt head. No finite location bound follows from this argument at or below $\eta^\star=1/2$, consistent with a flat block centered arbitrarily deep in rank.

Anti-flatness has a complementary global envelope. The finite-dimensional extremal-spectrum result and the dimension-independent supremum $27/256$ were established in Ref.~\cite{jasser2026journey}. Nonnegativity follows by viewing $X=\lambda_i$ as a random variable sampled with probability $\lambda_i$: then $\mathbb E[X]=P_2$, $\mathbb E[X^2]=P_3$, and $\mathcal F=\operatorname{Var}(X)\geq0$. For the upper bound, write $p=\lambda_0$ and $q=P_2$. Since every $\lambda_i\leq p$,
\begin{equation}
P_3=\sum_i\lambda_i^3\leq p\sum_i\lambda_i^2=pq,
\end{equation}
and hence $\mathcal F\leq q(p-q)$. The quadratic $q(p-q)$ is at most $p^2/4$, and the $\ell=0$ octave gives $p\leq\eta^\star$. These steps establish the concentration-dependent chain
\begin{equation}
0\leq\mathcal F\leq q(p-q)\leq\frac{p^2}{4}\leq\frac{(\eta^\star)^2}{4},\qquad \mathcal F\leq\frac{27}{256}.
\label{eq:s_af_bounds}
\end{equation}
To recover the dimension-independent constant in the last inequality, also use $q=P_2\geq p^2$. For $p\leq1/2$, unconstrained maximization over $q$ gives $q(p-q)\leq p^2/4\leq1/16$. For $p\geq1/2$, the parabola is decreasing throughout the allowed interval $q\geq p^2\geq p/2$, so
\begin{equation}
\mathcal F\leq p^2(p-p^2)=p^3(1-p)\leq\frac{27}{256},
\end{equation}
where the final supremum occurs at $p=3/4$. As established in Ref.~\cite{jasser2026journey}, this supremum is approached by the spectrum $\lambda_0=3/4$ followed by an increasingly diffuse tail of total weight $1/4$, for which $P_2\to p^2$, $P_3\to p^3$, and $\mathcal F\to27/256$.

\section{Head erosion and the ordering of spectral events}

This section gives the minimal analytic head--tail model for the clocks. It separates two microscopic ways to increase entropy and derives the order of the roughness, magic, and octave thresholds.

Write any ordered spectrum as $\lambda_0=1-\epsilon$ and $\lambda_i=\epsilon q_i$ for $i\geq1$, with $\sum_iq_i=1$. Here $\epsilon$ is the total tail weight and $\bm q$ is its normalized internal shape. Define the tail entropy $H_{\rm tail}=-\sum_iq_i\log_2q_i$ and its entropic rank $R_S=2^{H_{\rm tail}}$. Substitution into the von Neumann entropy separates the binary head--tail uncertainty from the entropy within the tail,
\begin{equation}
S=h_2(\epsilon)+\epsilon H_{\rm tail}
=h_2(\epsilon)+\epsilon\log_2R_S.
\end{equation}
Since $dh_2/d\epsilon=\log_2[(1-\epsilon)/\epsilon]$, differentiation gives the exact identity
\begin{equation}
dS=\log_2\frac{R_S(1-\epsilon)}{\epsilon}\,d\epsilon+\epsilon\,d\log_2R_S.
\label{eq:s_entropy_differential}
\end{equation}
The two terms separate limiting contributions to entropy growth rather than mutually exclusive dynamical stages. Head erosion transfers probability out of $\lambda_0$, so $\epsilon$ and hence the total tail weight necessarily increase. Tail-rank expansion has a different meaning: at fixed $\epsilon$, the normalized tail $\bm q$ spreads over more Schmidt modes and $R_S$ increases without further reducing the head. The two effects generally occur together in many-body dynamics, but Eq.~\eqref{eq:s_entropy_differential} distinguishes changes in total tail weight from changes in its internal width.

The first term is therefore head erosion at fixed normalized tail shape, and the second is tail-rank expansion at fixed total tail weight. On the physically relevant branch $R_S(1-\epsilon)>\epsilon$, the coefficient of $d\epsilon$ is positive. As long as the leading probability remains the dominant Schmidt-scale window, $\eta^\star=1-\epsilon$ and $d\eta^\star=-d\epsilon$. Pure head erosion then obeys
\begin{equation}
\left.\frac{d\eta^\star}{dS}\right|_{R_S}
=-\left[\log_2\frac{R_S(1-\epsilon)}{\epsilon}\right]^{-1}<0,
\end{equation}
whereas pure tail expansion has $dS=\epsilon\,d\log_2R_S>0$ and $d\eta^\star=0$. Thus, in the $(S,\eta^\star)$ plane, head erosion produces motion toward larger entropy and smaller concentration, while tail-rank expansion produces horizontal motion toward larger entropy at fixed concentration. A mixed trajectory interpolates between these limiting slopes: a pronounced negative slope indicates substantial transfer of weight out of the head, while a slope near zero indicates entropy growth dominated by broadening within the existing tail. The relevant contrast is therefore between negative and vanishing slopes.

\begin{figure}[tbp]
\includegraphics[width=0.60\textwidth]{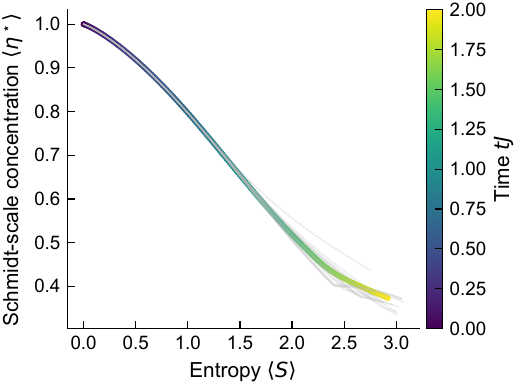}
\caption{Early-time concentration--entropy trajectories for periodic $L=14$ N\'eel quenches at $W=0.5$. Each thin gray curve follows one representative disorder realization parametrically from $tJ=0$ to $2$; the colored points follow the 256-realization mean, and the color bar gives time. The trajectory moves from the rank-one point $(S,\eta^\star)=(0,1)$ toward larger entropy and smaller concentration. Its Pearson coefficient is $r_{\rm traj}=-0.996$; the fixed-time realization correlation at $tJ=2$ is $-0.758$, showing that the anticorrelation is not only an artifact of using time as a common parameter.}
\label{fig:s_concentration_entropy}
\end{figure}

We use two nonredundant comparisons because successive times on one trajectory are correlated and are not independent samples. For the $L=14$, $W=0.5$ data in Fig.~\ref{fig:s_concentration_entropy}, the Pearson coefficient between the ensemble means $\langle S(t)\rangle$ and $\langle\eta^\star(t)\rangle$ across the sampled times is $r_{\rm traj}=-0.996$; this describes the direction of the mean dynamical path. At the single time $tJ=2$, the correlation across 256 disorder realizations is $r_{t=2}=-0.758$, showing that the negative association also exists between samples without treating different times as independent observations. Combining the $L=14$, $tJ=10^8$ endpoints from the six disorder ensembles $W=2$--8 gives $r=-0.942$; this last value primarily describes the separation of disorder regimes, since correlations evaluated within one fixed-$W$ ensemble need not have the same sign. These coefficients are descriptive statistics, while the sign mechanism follows from the differential identity above.

For a flat tail, $q_i=1/R$, the power sums are
\begin{equation}
P_2=(1-\epsilon)^2+\frac{\epsilon^2}{R},\qquad
P_3=(1-\epsilon)^3+\frac{\epsilon^3}{R^2}.
\end{equation}
Substituting these expressions into $\mathcal F=P_3-P_2^2$, and noting that the most massive tail octave contains $\lfloor R/2\rfloor+1$ equal tail entries, gives
\begin{align}
S&=h_2(\epsilon)+\epsilon\log_2R,\nonumber\\
\mathcal F&=\epsilon(1-\epsilon)\left[1-\left(1+\frac1R\right)\epsilon\right]^2,\nonumber\\
\eta^\star&=\max\left\{1-\epsilon,\frac{\lfloor R/2\rfloor+1}{R}\epsilon\right\}.
\label{eq:s_headtail}
\end{align}
The entropy formula follows directly from $S=h_2(\epsilon)+\epsilon H_{\rm tail}$ with $H_{\rm tail}=\log_2R$. For the octave result, the head window has mass $1-\epsilon$, while every tail-only window has mass equal to its number of included entries times $\epsilon/R$; the longest admissible tail window gives the second branch in Eq.~\eqref{eq:s_headtail}. The dominant-Schmidt-scale transport threshold is therefore
\begin{equation}
\epsilon_\ell(R)=\left[1+\frac{\lfloor R/2\rfloor+1}{R}\right]^{-1}\longrightarrow\frac23.
\end{equation}

To locate the roughness maximum, set $a=1+1/R$ and write $\mathcal F=\epsilon(1-\epsilon)(1-a\epsilon)^2$. Its derivative factorizes as
\begin{equation}
\frac{d\mathcal F}{d\epsilon}=-(a\epsilon-1)\left(4a\epsilon^2-(3a+2)\epsilon+1\right).
\end{equation}
The smaller root of the quadratic is the first interior maximum reached from the product state,
\begin{equation}
\epsilon_{\mathcal F}(R)=\frac{2+3a-\sqrt{9a^2-4a+4}}{8a},\qquad a=1+\frac1R,
\label{eq:s_af_threshold}
\end{equation}
which tends to $1/4$ as $R\to\infty$; the larger root is the later stationary point and $\epsilon=1/a$ is a zero of $\mathcal F$.

When $R+1=2^\nu$, a dyadic prefix of size $m=2^k$ has fidelity
\begin{equation}
F_m=\frac{\left[\sqrt{1-\epsilon}+(m-1)\sqrt{\epsilon/R}\right]^2}{m}
=B^2m+2B(A-B)+\frac{(A-B)^2}{m},
\end{equation}
where $A=\sqrt{1-\epsilon}$ and $B=\sqrt{\epsilon/R}$. Treating $m$ as continuous, $d^2F_m/dm^2=2(A-B)^2/m^3\geq0$. A convex function on $1\leq m\leq R+1$ is maximized at an endpoint; restricting to dyadic $m$ therefore leaves only the rank-one and full-rank stabilizer sectors,
\begin{equation}
F_{\rm NL}=\max\left\{1-\epsilon,\frac{\left(\sqrt{1-\epsilon}+\sqrt{R\epsilon}\right)^2}{R+1}\right\}.
\label{eq:s_headtail_magic}
\end{equation}
The two branches are equal when $1-\epsilon=(\sqrt{1-\epsilon}+\sqrt{R\epsilon})^2/(R+1)$. Taking the positive square root, isolating the ratio $\sqrt{\epsilon/(1-\epsilon)}$, and squaring gives
\begin{equation}
\epsilon_D(R)=\frac{(\sqrt{R+1}-1)^2}{R+(\sqrt{R+1}-1)^2}\longrightarrow\frac12.
\label{eq:s_magic_threshold}
\end{equation}
For the $R=63$ spectrum plotted in Fig.~3(a) of the Letter, the three thresholds are $\epsilon_{\mathcal F}=0.24738$, $\epsilon_D=7/16$, and $\epsilon_\ell=63/95$; in particular, $\epsilon_D(R)=1/2-1/(2\sqrt R)+O(R^{-1})$ approaches its limit relatively slowly. The dominant Schmidt scale moves when the second branch of Eq.~\eqref{eq:s_headtail} exceeds $1-\epsilon$, giving the finite-$R$ threshold above and $\epsilon_\ell\to2/3$. The binary spectrum $R=1$ is special: Eqs.~\eqref{eq:s_af_threshold} and~\eqref{eq:s_magic_threshold} both give $\epsilon=(2-\sqrt2)/4$, so roughness and nonlocal magic peak at the same spectrum, while the ordered rank-two spectrum remains head dominated. Increasing $R$ separates these two extrema and creates tail octaves capable of becoming dominant; for a broad tail, $\epsilon_{\mathcal F}<\epsilon_D<\epsilon_\ell$. Thus one finite-$R$ family connects the early binary coincidence to the later many-mode ordering. On the rank-one fidelity branch, $D_{\min}^{\rm NL}=-\log_2(1-\epsilon)$ increases with head erosion; after the optimizer switch, the full-rank fidelity increases toward one and the magic falls toward the dyadic-flat endpoint. These locations become clock times only after a microscopic evolution specifies $\epsilon(t)$ and $R(t)$. In particular, $dS/dt$ depends on both $\dot\epsilon$ and $d\log_2R_S/dt$, as supplied by the active-bond dynamics below.

\section{Active exchange bonds across an entanglement cut}

The head--tail model specifies thresholds in spectrum space without assigning them times. This section supplies a microscopic trajectory near the beginning of an XXZ quench, including the entropy-rate peak that spectral geometry alone cannot fix, and explains the dependence on bipartition cuts and initial states.

We first fix the physical cut geometry convention. A nearest-neighbor open half chain has one bond connecting $A$ and $B$, while a periodic half chain or an interval inside an open chain has two. Only such cut-crossing bonds can be active exchange bonds, and a cut bond is active for a product state only when its endpoint spins are antiparallel; bonds lying entirely inside $A$ or $B$ are never counted. Thus the numerical protocols have $n_{\rm active}\in\{0,1\}$ for one cut and $n_{\rm active}\in\{0,1,2\}$ for two cuts. We retain a general integer $n$ in the factorized formulas solely to state the independent-bond tensor-product identity compactly; its XXZ applications here use only $n=1$ and $2$. The arbitrary number of cross-cut pairs in the l-bit model is introduced separately in Sec.~6.

For one antiparallel spin pair crossing the cut, the exchange term preserves the two-dimensional active subspace $\{|\uparrow\downarrow\rangle,|\downarrow\uparrow\rangle\}$. After removing a common energy, the restricted Hamiltonian is
\begin{equation}
H_{\rm cut}=\frac12(J\sigma^x+\delta\sigma^z),\qquad \delta=h_L-h_R.
\label{eq:s_cut_hamiltonian}
\end{equation}
Let $\Omega=\sqrt{J^2+\delta^2}$. Since $H_{\rm cut}^2=\Omega^2\mathbb I/4$, its exponential is
\begin{equation}
e^{-itH_{\rm cut}}=\cos\frac{\Omega t}{2}\,\mathbb I-\frac{2i}{\Omega}\sin\frac{\Omega t}{2}\,H_{\rm cut}.
\end{equation}
Starting from $|\uparrow\downarrow\rangle$, the squared amplitude transferred to $|\downarrow\uparrow\rangle$ and the resulting two Schmidt probabilities are therefore
\begin{equation}
q_\delta(t)=\frac{J^2}{J^2+\delta^2}\sin^2\left(\frac{\sqrt{J^2+\delta^2}\,t}{2}\right),\qquad \bm\lambda=(1-q_\delta,q_\delta).
\label{eq:s_binary_spectrum}
\end{equation}
At resonance, write $x=Jt$ and $q=\sin^2(x/2)$. The entropy rate is
\begin{equation}
\frac{dS}{dt}=\frac{J\sin x}{2}\log_2\frac{1-q}{q}.
\end{equation}
Its first maximum solves $\cos x\ln[\cot^2(x/2)]=2$, giving $t_{\dot S}^{\rm bond}J=0.585282$. For the binary spectrum,
\begin{equation}
\mathcal F=q(1-q)(1-2q)^2,\qquad
\frac{d\mathcal F}{dq}=-(2q-1)(8q^2-8q+1).
\end{equation}
The first nontrivial root is $q=(2-\sqrt2)/4$, hence $t_{\mathcal F}^{\rm bond}J=\pi/4$. When $h_L$ and $h_R$ are independent uniform variables on $[-W,W]$, their difference has triangular density $p_W(\delta)=(2W-|\delta|)/(4W^2)$ on $|\delta|\leq2W$. Disorder-averaged bond predictions are obtained by integrating Eq.~\eqref{eq:s_binary_spectrum} against this normalized density.

For the independent-bond normal form, retain only the $n$ active cut bonds and remove all couplings between their endpoint pairs and within either half. Each retained bond evolves only in $\{|\uparrow\downarrow\rangle,|\downarrow\uparrow\rangle\}$ and creates the same binary Schmidt spectrum $(1-q,q)$. The total state is a tensor product of these bond states, so every Schmidt probability is a product of $n$ factors chosen from $q$ and $1-q$. Additivity of Shannon entropy and factorization of power sums then give
\begin{equation}
S_{\rm bond}^{(n)}=nh_2(q),\qquad P_m^{(n)}=[p_m(q)]^n,\qquad p_m(q)=(1-q)^m+q^m,\qquad \mathcal F^{(n)}=P_3^{(n)}-[P_2^{(n)}]^2.
\label{eq:s_multibond}
\end{equation}
Because $S_{\rm bond}^{(n)}=n h_2(q)$, multiplication by $n$ changes the height of the entropy-rate curve but not its maximizing time. Anti-flatness is nonlinear in the factorized moments, so its stationarity condition becomes
\begin{equation}
(2q-1)\left[3p_3(q)^{n-1}-4p_2(q)^{2n-1}\right]=0,
\end{equation}
apart from the positive factor $n$. Solving the first interior root shifts the roughness clock from $0.785398/J$ for one active bond to $0.652620/J$ for two. The one-bond case is exactly the binary endpoint $R=1$ of the head--tail family. For two equivalent active bonds and $0\leq q\leq1/2$, the ordered spectrum is $((1-q)^2,q(1-q),q(1-q),q^2)$. Its dyadic-prefix fidelities obey
\begin{equation}
F_1=(1-q)^2,\qquad F_2=\frac{1-q}{2}\left(\sqrt{1-q}+\sqrt q\right)^2,\qquad F_4=\frac14\left(\sqrt{1-q}+\sqrt q\right)^4,
\end{equation}
with $F_2^2=F_1F_4$. The optimal sector therefore switches directly from rank one to rank four when all three coincide, at $q=(2-\sqrt2)/4$. Hence the two-bond magic clock remains $t_D^{\rm bond}J=\pi/4=0.785398$, while its roughness clock advances to $0.652620$. In the general head--tail decomposition, the two-bond spectrum has $\epsilon=2q-q^2$ and normalized tail $[q(1-q),q(1-q),q^2]/(2q-q^2)$. It is therefore a microscopic nonflat-tail trajectory rather than a second use of the flat-tail ansatz. Its head window has mass $\eta_0=(1-q)^2$, and the first tail window has mass $\eta_1=2q(1-q)$. Their equality gives $q=1/3$, so the first octave switch occurs at
\begin{equation}
t_\ell J=2\arcsin(1/\sqrt3)=1.230959.
\label{eq:s_two_bond_unpin}
\end{equation}
Before this crossing, the dominant concentration is the decreasing head branch $\eta_0=(1-q)^2$; afterward, the first transported window $\eta_1=2q(1-q)$ is larger. Their upper envelope $\eta^\star=\max\{\eta_0,\eta_1\}$ therefore has a cusp-shaped minimum $\eta^\star=4/9$ at the clean crossing. On later coherent branches, each binary factor must be reordered by replacing $q$ with $\min(q,1-q)$ before the product spectrum is sorted. The vertical line in Fig.~\ref{fig:s_unpinning}(a),(c) is this parameter-free clean reference; detuning $\delta$ and disorder averaging round the cusp, while the further delay of the numerical transported fraction measures dressing by the intra-half many-body dynamics.

For a uniformly sampled half-filled product state, let $N=L/2$ be the number of up spins. A specified bond is active when it contains one up and one down spin. There are two orientations and $\binom{L-2}{N-1}$ compatible configurations of the remaining sites, hence
\begin{equation}
P(\text{one specified bond active})=
\frac{2\binom{L-2}{N-1}}{\binom{L}{N}}=\frac{L}{2(L-1)}.
\end{equation}
For two disjoint periodic cut bonds, direct counting gives
\begin{align}
\pi_2&=\frac{4\binom{L-4}{N-2}}{\binom{L}{N}},\\
\pi_1&=\frac{4\left[\binom{L-4}{N-1}+\binom{L-4}{N-3}\right]}{\binom{L}{N}},\\
\pi_0&=\frac{\binom{L-4}{N}+2\binom{L-4}{N-2}+\binom{L-4}{N-4}}{\binom{L}{N}}.
\end{align}
At $L=14$ these probabilities are $0.2937$, $0.4895$, and $0.2168$ for two, one, and zero active exchange bonds, consistent with the measured fractions $0.3057$, $0.4971$, and $0.1973$. This active-bond count explains both the protocol mixture and the ordering of the early clocks. To isolate the residual transport delay, we keep the two cross-cut XXZ bonds at strength $J$ and multiply every bond wholly within either half by $g$. The interpolation from $g=0$ (two isolated cut bonds) to $g=1$ (the complete periodic chain) is shown in Fig.~\ref{fig:s_bond_controls}(c). At $g=0$, the numerical clocks $(t_{\dot S}^\star,t_{\mathcal F}^\star,t_D^\star)J=(0.58,0.65,0.79)$ agree with the independent-bond predictions $(0.585282,0.652620,0.785398)$.

\section{Interacting l-bit dephasing and dominant-Schmidt-scale transport}

This section analyzes the dynamics of Anderson localization and an l-bit effective model of MBL. The analysis proceeds from the terms that change the Schmidt spectrum, to one reduced l-bit mode, to the many-pair spectrum, and finally to the logarithmic activation of increasingly distant pairs. The single-mode calculation supplies a local clock benchmark; the many-pair calculation explains pinning and transport; and the final comparison states exactly which predictions are tested numerically. We use the diagonal l-bit Hamiltonian
\begin{equation}
H_{\rm lbit}=\sum_i h_i\tau_i^z+\sum_{i<j}J_{ij}\tau_i^z\tau_j^z+\cdots,\qquad |J_{ij}|\sim J_0e^{-|i-j|/\xi}.
\label{eq:s_lbit}
\end{equation}
Here $\tau_i^z$ is a quasi-local Pauli integral of motion, $h_i$ is its local field, and $J_{ij}$ is the interaction between l-bits $i$ and $j$; $J_0$ sets the interaction scale and $\xi$ is the localization length. The omitted terms denote higher-body products of l-bit integrals of motion. The initial state is essential in this diagonal representation. Any product state in the $\tau^z$ basis is an exact eigenstate of Eq.~\eqref{eq:s_lbit} and has trivial dynamics, whereas $|+\rangle_\tau^{\otimes L}$ has maximal coherence across l-bit configurations and exposes the full dephasing mechanism. A physical-spin N\'eel state is not exactly a $\tau^z$ product because the l-bits are quasi-locally dressed, but it approaches that limit as localization strengthens. The l-bit calculation below is therefore a solvable mechanism limit for a coherent initial state, not a state-independent prediction for every MBL quench.

\textit{Cross-cut origin.---}Terms supported entirely within $A$ or $B$ generate local unitaries and preserve the Schmidt spectrum, so only cross-cut interactions need to be retained. Label a $\tau^z$ configuration in $A$ by $\bm\sigma$ and one in $B$ by $\bm\mu$, with entries $\pm1$. The initial $\tau^x$ product state has equal amplitude for every pair $(\bm\sigma,\bm\mu)$. Evolution multiplies that amplitude by the phase generated by $E_A(\bm\sigma)+E_B(\bm\mu)+\sum_{i\in A,j\in B}J_{ij}\sigma_i\mu_j$. In the partial trace, the phase $E_B(\bm\mu)$ cancels between bra and ket. Summing each traced spin $\mu_j=\pm1$ independently uses $\frac12\sum_{\mu_j=\pm1}e^{-it\mu_jx}=\cos(tx)$ and gives
\begin{equation}
\rho_A(\bm\sigma,\bm\sigma';t)=2^{-|A|}e^{-it[E_A(\bm\sigma)-E_A(\bm\sigma')]}
\prod_{j\in B}\cos\left[t\sum_{i\in A}J_{ij}(\sigma_i-\sigma_i')\right].
\label{eq:s_lbit_rho}
\end{equation}
Equation~\eqref{eq:s_lbit_rho} is the exact bridge from unitary l-bit evolution to the Schmidt spectrum. Its diagonal entries are unchanged, while each off-diagonal element is multiplied by a product of cosines generated only by interactions crossing the bipartition. The prefactor containing $E_A$ is a diagonal unitary conjugation and cannot change the eigenvalues of $\rho_A$. Thus cross-cut interactions alone determine the time-dependent Schmidt weights. Higher-body l-bit terms replace the cosine product by analogous characteristic functions without changing this separation between intra-side phases and cross-cut dephasing.

\textit{Single-mode benchmark.---}A single cross-cut l-bit pair provides the elementary spectral building block. For one l-bit in $A$ coupled to one l-bit in $B$, setting $\sigma=-\sigma'$ in Eq.~\eqref{eq:s_lbit_rho} gives the coherence factor $\Gamma(t)=\cos(2J_{ij}t)$. After removing the local phase by a basis rotation,
\begin{equation}
\rho_A(t)=\frac12\begin{pmatrix}1&\Gamma(t)\\ \Gamma(t)&1\end{pmatrix}
=\frac{\mathbb I+\Gamma(t)\sigma^x}{2},
\label{eq:s_one_pair_rho}
\end{equation}
with eigenvalues $\lambda_\pm=(1\pm\Gamma)/2$ on the first decay branch where $\Gamma\geq0$. Equivalently, the reduced evolution is the dephasing map $\mathcal D_\Gamma(\rho)=[(1+\Gamma)\rho+(1-\Gamma)\tau^z\rho\tau^z]/2$: this is an exact description obtained after tracing the partner l-bit, not an externally imposed noise channel. Substituting the two eigenvalues into the spectral observables gives $P_2=(1+\Gamma^2)/2$ and $P_3=(1+3\Gamma^2)/4$, hence
\begin{equation}
\mathcal F=\frac{\Gamma^2(1-\Gamma^2)}{4},\qquad F_{\rm NL}=\max\left\{\frac{1+\Gamma}{2},\frac{1+\sqrt{1-\Gamma^2}}{2}\right\}.
\label{eq:s_one_dephasing}
\end{equation}
The same eigenvalues give $S=h_2[(1-\Gamma)/2]$, so entropy rises from zero to one bit as coherence decays from $\Gamma=1$ to zero. A binary spectrum cannot transport its dominant Schmidt scale: its two allowed octave weights are $\lambda_+$ and $\lambda_-$, so $\eta^\star=\lambda_+$ and $\ell^\star=0$ throughout this branch. The anti-flatness derivative is proportional to $\Gamma(1-2\Gamma^2)$, so its interior maximum occurs at $\Gamma=1/\sqrt2$. The two fidelity branches in Eq.~\eqref{eq:s_one_dephasing} are equal at the same value, proving that the roughness maximum and the nonlocal-magic optimizer switch coincide for one dephasing mode. Equation~\eqref{eq:s_one_pair_rho} therefore fixes the common local roughness--magic threshold and shows why additional l-bit pairs are necessary for dominant-Schmidt-scale transport.

This binary mode also gives a useful local clock benchmark. For one tagged l-bit in $A$ coupled to many traced l-bits in $B$, the reduced spectrum remains rank two and its coherence is $\Gamma(t)=\prod_j\cos(2J_{ij}t)$. When no single weak coupling dominates, a second-cumulant approximation gives $\Gamma(t)\simeq e^{-\gamma t^2}$ with $2\gamma\equiv4\sum_{j\in B}J_{ij}^2$. Writing $x=\sqrt\gamma t$, the entropy-production rate is proportional to $x e^{-x^2}\ln[(1+e^{-x^2})/(1-e^{-x^2})]$. Maximizing it and imposing the roughness condition $\Gamma=1/\sqrt2$ give
\begin{equation}
x\equiv\sqrt{\gamma}\,t,\qquad x_{\dot S}=0.364251,\qquad x_{\mathcal F}=\sqrt{\frac{\ln2}{2}}=0.588705,\qquad \frac{t_{\mathcal F}^\star}{t_{\dot S}^\star}=1.61621.
\label{eq:s_single_mode_clock}
\end{equation}
This numerical ratio is close to the weak-disorder open-boundary XXZ value $0.88/0.53\simeq1.66$, but the agreement is phenomenological: Eq.~\eqref{eq:s_single_mode_clock} assumes Gaussian l-bit dephasing, whereas the early XXZ clocks arise from coherent exchange and many-body dressing. It is not a prediction for the global many-pair l-bit dynamics.

\textit{Many-pair spectrum.---}The full bipartition is controlled by the product of many binary Schmidt factors. These cross-cut l-bit pairs are distinct from the one or two nearest-neighbor exchange bonds in the preceding section. For independent pairs with ordered weights $(1-q_a,q_a)$, where $0\leq q_a\leq1/2$, Shannon entropy is additive, each power sum factorizes, and the largest probability selects the larger weight from every pair:
\begin{equation}
S=\sum_a h_2(q_a),\qquad P_n=\prod_a[(1-q_a)^n+q_a^n],\qquad \lambda_0=\prod_a(1-q_a).
\label{eq:s_lbit_product}
\end{equation}

\textit{Pinning versus transport.---}Equation~\eqref{eq:s_lbit_product}, rather than the single-mode ratio, controls the global observables. Its first exact consequence is head pinning: if $\lambda_0\geq1/2$, the entire tail has mass $1-\lambda_0\leq\lambda_0$, so no tail-only octave can overtake the head. The same head weight also bounds the optimal dyadic fidelity. A dyadic prefix of size $m\geq2$ obeys Cauchy--Schwarz,
\begin{equation}
F_m=\frac1m\left(\sqrt{\lambda_0}+\sum_{i=1}^{m-1}\sqrt{\lambda_i}\right)^2
\leq\frac1m\left(\sqrt{\lambda_0}+\sqrt{(m-1)(1-\lambda_0)}\right)^2.
\end{equation}
The right-hand side does not exceed the rank-one fidelity $F_1=\lambda_0$ whenever $\lambda_0\geq(\sqrt m+1)/(2\sqrt m)$. This threshold is largest at $m=2$, yielding
\begin{equation}
\lambda_0\geq\frac{2+\sqrt2}{4}\Rightarrow D_{\min}^{\rm NL}=-\log_2\lambda_0,\qquad \lambda_0\geq\frac12\Rightarrow\eta^\star=\lambda_0,\ \ell^\star=0.
\label{eq:s_pinning}
\end{equation}
The many-pair pinning result is the central link to dominant-Schmidt-scale transport. Entropy can grow through many weak cross-cut interactions while $\lambda_0\geq1/2$ keeps the dominant octave pinned to the head. The magic condition is stronger because $(2+\sqrt2)/4>1/2$: the rank-one stabilizer sector can cease to be optimal before the dominant Schmidt scale moves.

The converse of the pinning criterion is not automatic: $\lambda_0<1/2$ permits dominant-Schmidt-scale transport but does not by itself guarantee it. The independent-pair spectrum nevertheless gives a constructive transport mechanism. For two equally dephased pairs, the ordered spectrum is $((1-q)^2,q(1-q),q(1-q),q^2)$; the first tail octave overtakes the head when $2q(1-q)>(1-q)^2$, or $q>1/3$. More generally, if $N_{\rm act}$ pairs are fully dephased, $q_a=1/2$, their spectrum is flat of rank $2^{N_{\rm act}}$ and
\begin{equation}
\eta^\star=\frac12,\qquad \ell^\star=2^{N_{\rm act}-1}-1,\qquad u^\star=N_{\rm act}-1.
\label{eq:s_lbit_flat_front}
\end{equation}
Thus each additional fully dephased pair doubles the occupied Schmidt rank and advances the dominant logarithmic scale by one. This flat-spectrum limit proves that many-pair dephasing can transport the dominant scale; it does not imply monotone motion in every realization, because unequal couplings and random phases can delay or reverse individual octave switches.

\textit{Logarithmic activation.---}The spatial hierarchy of l-bit couplings supplies the many-body time dependence. Interactions at distance $r$ dephase when $|J(r)|t\sim1$. With $|J(r)|\sim J_0e^{-r/\xi}$, solving this condition gives $r(t)\simeq\xi\ln(J_0t)$. If $\rho_\times$ statistically independent cross-cut pairs are available per unit distance, their number is therefore
\begin{equation}
N_{\rm act}(t)\simeq\rho_\times\xi\ln(J_0t).
\end{equation}
The logarithmically growing active set converts the product formulas into robust long-time trends without requiring a specific phase distribution. Once active pairs have dephased, suppose each contributes a finite mean entropy $\overline{s}>0$, a finite typical logarithmic head cost $a_\eta>0$, and finite logarithmic moment costs $b_n>0$. Additivity of entropy and factorization of the head weight and power sums then give
\begin{equation}
S\simeq\overline{s}N_{\rm act}\propto\ln t,\qquad \eta^\star_{\rm typ}=\lambda_{0,\rm typ}\sim e^{-a_\eta N_{\rm act}}\sim t^{-\alpha_\eta},\qquad P_{n,\rm typ}\sim e^{-b_nN_{\rm act}}\sim t^{-\alpha_n},
\label{eq:s_lbit_scaling}
\end{equation}
on the head-dominated branch and before finite-size saturation, with positive model-dependent exponents $\alpha_\eta$ and $\alpha_n$. Here the subscript ${\rm typ}$ denotes a typical value, equivalently the exponential of the disorder-averaged logarithm in this multiplicative approximation. Because $\mathcal F=P_3-P_2^2$, its post-barrier decay is likewise algebraic in the independent-pair picture. Together, Eqs.~\eqref{eq:s_lbit_flat_front} and \eqref{eq:s_lbit_scaling} bracket the scale dynamics between an initially pinned head and an ideal moving front with $u^\star\simeq N_{\rm act}-1\propto\ln t$. The central prediction is therefore qualitative but falsifiable: exponentially weak cross-cut interactions produce logarithmic entropy growth, algebraic head and moment decay, and statistically sustained dominant-Schmidt-scale transport, whereas noninteracting localization cannot sustain this spectral front. Correlated l-bits, broad coupling prefactors, higher-body terms, and quasi-local basis dressing change the coefficients and the detailed motion without altering this mechanism-level distinction~\cite{bardarson2012unbounded,serbyn2013universal,serbyn2013local,huse2014phenomenology,abanin2019many}.

\textit{Relation to numerics.---}The numerical control in Fig.~\ref{fig:s_localization} tests these functional predictions. Its l-bit Hamiltonian contains every pair interaction $J_{ij}$, so different interactions share l-bits and generate correlated Schmidt factors. Nevertheless, it shows the predicted sequence: over $1\leq tJ\leq100$, $d\langle S\rangle/d\ln t=0.792$; concentration falls; anti-flatness reaches a transient maximum and then decays; and the dominant Schmidt scale eventually leaves the spectral head. At $tJ=10^8$, the l-bit ensemble has $\langle S\rangle=5.266$, $\langle\eta^\star\rangle=0.307$, $\langle u^\star\rangle=3.83$, and unit transported fraction, whereas the Anderson ensemble remains head-pinned with $\langle S\rangle=0.296$, $\langle\eta^\star\rangle=0.923$, and $\langle u^\star\rangle=0$. The agreement is therefore at the mechanism level: interactions sustain a logarithmic spectral front and dominant-Schmidt-scale transport, while the detailed exponents remain model dependent.

\section{Numerical evidence and mechanism controls}

This section gathers the extended data behind the three main figures.
\subsection{Early-time protocols and dominant-Schmidt-scale transport}

\begin{figure}[tbp]
\includegraphics[width=0.98\textwidth]{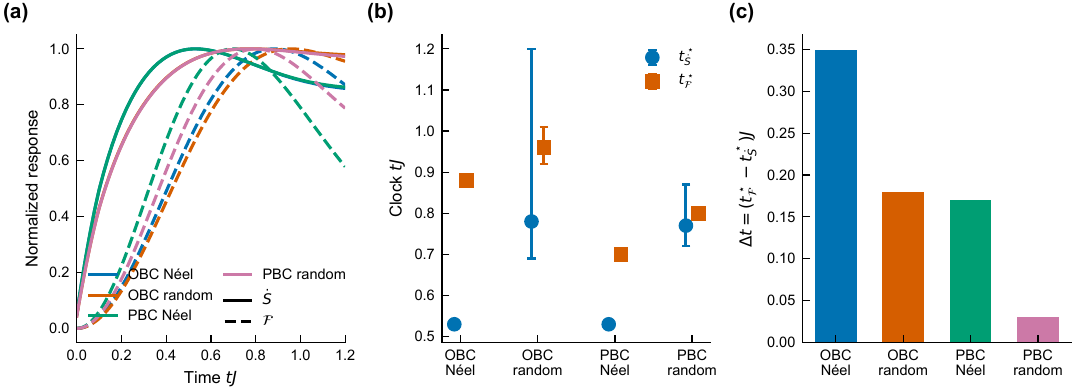}
\caption{Protocol dependence of the early entropy and roughness clocks at $L=14$ and $W=0.5$. (a) Normalized entropy-production rate $\dot S/\dot S_{\max}$ (solid) and anti-flatness $\mathcal F/\mathcal F_{\max}$ (dashed) versus time for N\'eel and random half-filled product states with open or periodic boundary conditions; color identifies the four protocols. (b) Extracted peak times $t_{\dot S}^\star$ (circles) and $t_{\mathcal F}^\star$ (squares); error bars are 95\% trajectory-bootstrap intervals. (c) Separation $t_{\mathcal F}^\star-t_{\dot S}^\star$ for the same protocols. The open-boundary N\'eel, open-boundary random, periodic-boundary N\'eel, and periodic-boundary random ensembles contain 128, 512, 512, and 1024 disorder realizations, respectively. Random initial cut patterns delay entropy production, while the second entanglement boundary under periodic boundary conditions advances roughening, producing the near coincidence for random product states.}
\label{fig:s_early_protocols}
\end{figure}

The first control determines why changing from the two active cut bonds of the main-text PBC N\'eel protocol to a single active OBC cut bond nearly merges the two middle clocks, and why earlier magic-barrier simulations found close entropy and anti-flatness clocks~\cite{zhang2026revealing}. At $L=14$ and $W=0.5$, the pairs $(t_{\dot S}^\star,t_{\mathcal F}^\star)$ are $(0.53,0.88)$ for a N\'eel product state with open boundary conditions, $(0.78,0.96)$ for random product states with open boundary conditions, $(0.53,0.70)$ for a N\'eel product state with periodic boundary conditions, and $(0.77,0.80)$ for random product states with periodic boundary conditions. The magic clock remains near $t_D^\star J=0.87$ for both N\'eel protocols, so the PBC pair $(t_{\mathcal F}^\star,t_D^\star)=(0.70,0.87)$ becomes $(0.88,0.87)$ under OBC\@. At fixed boundary condition, randomizing the product state delays $t_{\dot S}^\star$ by $0.24$--$0.25$ and delays $t_{\mathcal F}^\star$ by only $0.08$--$0.10$. At fixed initial-state class, adding the second entanglement boundary leaves $t_{\dot S}^\star$ unchanged within $0.01$ but advances $t_{\mathcal F}^\star$ by $0.16$--$0.18$. The two effects therefore act primarily on different clocks and jointly produce the near coincidence for periodic random product states [Fig.~\ref{fig:s_early_protocols}]. The previously reported near coincidence is therefore a property of that protocol rather than a universal relation between the two clocks.

The one-bond result has a simple spectral origin. The binary Schmidt spectrum $(1-q,q)$ generated by one active bond reaches its anti-flatness maximum and switches its optimal dyadic-flat stabilizer sector at the same value $q=(2-\sqrt2)/4$, locking $t_{\mathcal F}^\star=t_D^\star$ in the isolated limit. Two simultaneous active bonds produce a product spectrum: anti-flatness responds to both factors and peaks earlier, while the magic-sector switch remains near the single-bond value, yielding the ordered middle pair in the PBC N\'eel protocol. Detuning, many-body dressing, and discrete sector changes account for the residual splittings.

\begin{figure}[tbp]
\includegraphics[width=0.82\textwidth]{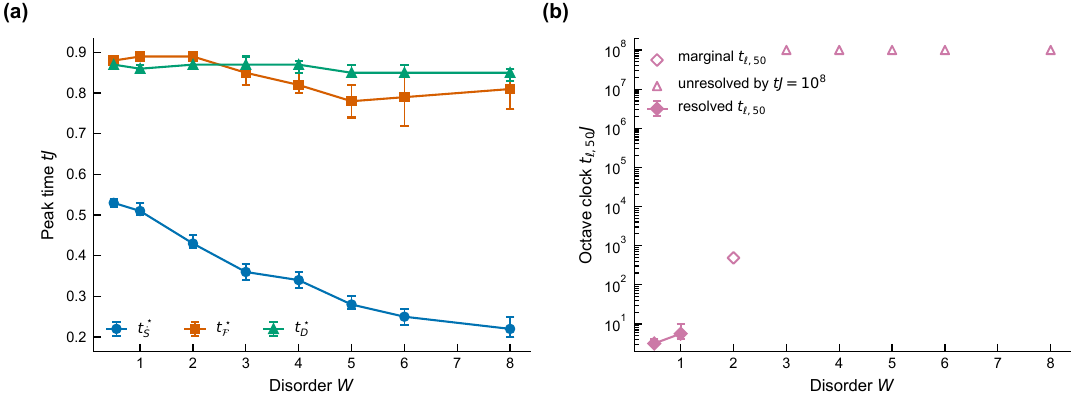}
\caption{Disorder dependence of all four open-chain N\'eel clocks at $L=14$. (a) The entropy-rate, anti-flatness, and exact-nonlocal-magic peak times from 128 realizations; error bars are 95\% trajectory-bootstrap intervals. (b) The majority-transport time $t_{\ell,50}$. Filled diamonds mark robust crossings at $W=0.5$ and $1$; the open diamond at $W=2$ marks a marginal crossing, for which 96\% of bootstrap ensembles cross and the conditional interval spans nearly three decades. Open triangles mark ensembles that remain below 50\% through $tJ=10^8$. Panel (b) uses 128 realizations for $W=1$--$4$ and 32 for $W=0.5,5,6,8$. Disorder advances the entropy-rate clock strongly, shifts the two spectral-shape peaks only weakly, and suppresses dominant-Schmidt-scale transport much more strongly.}
\label{fig:s_clock_disorder}
\end{figure}

The corresponding open-chain disorder scan is shown in Fig.~\ref{fig:s_clock_disorder}. For N\'eel product-state quenches with open boundary conditions, $t_{\dot S}^\star$ decreases monotonically from $0.53$ at $W=0.5$ to $0.20$--$0.22$ at $W=8$, whereas $t_{\mathcal F}^\star$ remains in $0.78$--$0.90$ and $t_D^\star$ in $0.85$--$0.88$. The maximum spread among $L=10,12,14$ is $0.02$, $0.03$, and $0.02$, respectively, so the trends are disorder driven rather than finite-size drift. Stronger fields suppress the delayed many-body contribution to the entropy-growth rate but move the two local spectral-shape barriers only weakly; consequently the separation between entropy production and spectral roughening grows with $W$. Throughout the OBC scan the middle clocks remain close and can exchange their point-estimate order, consistent with the single-active-bond locking mechanism. Both peaks diagnose the initial low-rank spectrum, whereas $u^\star$ diagnoses the later relocation of its dominant scale.

The fourth clock cannot be continued across the same disorder range as an ordinary peak time. Operationally, $t_{\ell,50}=\inf\{t:P(\ell^\star>0,t)\geq1/2\}$, with $\inf\varnothing=+\infty$. The 50\% criterion is robustly reached at $t_{\ell,50}J=3.16$ for $W=0.5$ and $5.62$ for $W=1$. At $W=2$ the point estimate is $486.97$, but the ensemble reaches only a maximum transported fraction of $0.547$; 4\% of bootstrap ensembles never cross, and the conditional interval spans nearly three decades. We therefore label this crossing marginal. For $W=3$ and $4$, the maximum transported fractions are only $0.273$ and $0.172$, respectively, and $W=5,6,8$ remain still lower, so all are unresolved through $tJ=10^8$. In these finite systems the long-time transported fraction may remain below one half, in which case the majority-transport time is infinite rather than merely beyond the observation window. This is the localized-phase alternative represented by the upper-edge markers in main Fig.~2(d): at finite observation time they are lower bounds, while physically they may signal an event that never occurs. Disorder therefore leaves the initial low-rank spectral reorganization operative while suppressing the collective redistribution needed for dominant-Schmidt-scale transport.

\begin{figure}[tbp]
\includegraphics[width=0.98\textwidth]{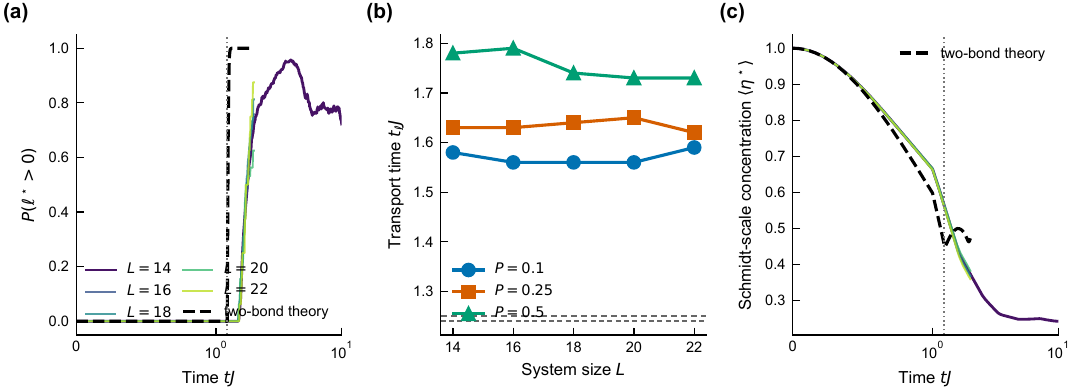}
\caption{Finite-size onset of dominant-Schmidt-scale transport for N\'eel product-state quenches with periodic boundary conditions at $W=0.5$. (a) Transported fraction $P(\ell^\star>0)$, the probability that the dominant Schmidt scale has left the spectral head. The $L=14$ curve continues through $tJ=10$; the $L=16,18,20,22$ curves resolve the common onset through $tJ=2$. The black dashed curve is the disorder-averaged independent-two-bond unpinning prediction and the gray vertical dotted line is the local clean resonant threshold prediction $tJ=2\arcsin(1/\sqrt3)$. (b) First crossing times at transported fractions $P=0.10$, $0.25$, and $0.50$ versus system size. The two-bond predictions for $P=0.10$ and $0.25$ both equal $tJ=1.24$ on the $0.01$ time grid and therefore share the lower horizontal dashed line; the upper dashed line at $1.25$ is the $P=0.50$ prediction. (c) Mean Schmidt-scale concentration for the same ensembles, including the extended $L=14$ trajectory, compared with the reordered two-bond result and clean threshold. Its dip is the rounded cusp where the decreasing head branch and increasing transported-window branch exchange dominance; disorder smears this feature, and the later numerical onset quantifies positive many-body dressing beyond the two-bond baseline. The five sizes use 256, 128, 64, 32, and 16 disorder realizations, respectively.}
\label{fig:s_unpinning}
\end{figure}

Figure~\ref{fig:s_unpinning} follows the onset of dominant-Schmidt-scale transport for N\'eel product-state quenches with periodic boundary conditions through $L=22$. The 10\%, 25\%, and 50\% transport times lie in $1.56$--$1.59$, $1.62$--$1.65$, and $1.73$--$1.79$, respectively. Their weak size dependence establishes the fourth clock as an early-time many-body feature. The clean two-bond estimate in Eq.~\eqref{eq:s_two_bond_unpin} captures the ordering and scale; intra-half dressing accounts for the remaining delay.

\begin{figure}[tbp]
\includegraphics[width=0.98\textwidth]{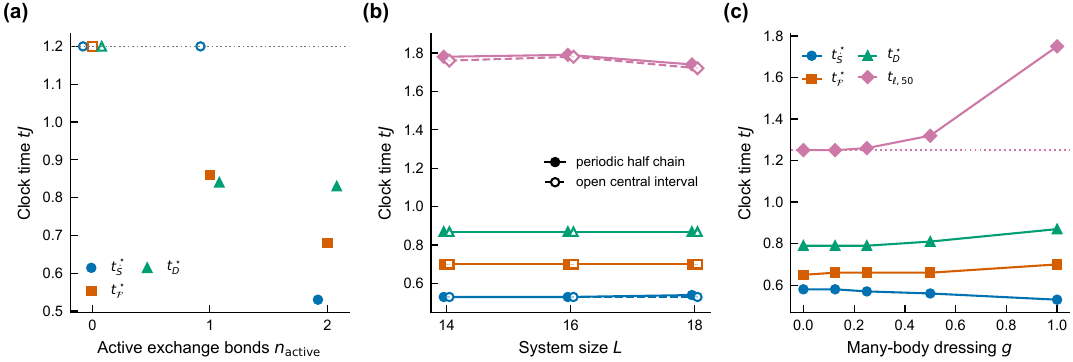}
\caption{Mechanism controls for the early-time hierarchy at $W=0.5$. (a) Entropy-rate, anti-flatness, and nonlocal-magic peak times for 1024 periodic $L=14$ random half-filled product-state trajectories, conditioned on the number $n_{\rm active}=0,1,2$ of initially active exchange bonds across the two entanglement cuts. Open symbols sit at the edge $tJ=1.2$ when a peak is not resolved inside the simulated window. (b) Absolute values of all four clocks for a periodic half chain (filled symbols) and an open-chain central interval (open symbols), both with two entanglement cuts. The paired values nearly overlap for $L=14,16,18$, directly showing that cut multiplicity rather than the global boundary condition controls the response. (c) Four clocks when all intra-half XXZ couplings are multiplied by the dressing strength $g$ while both cut couplings are held fixed, using 256 disorder realizations at each $g$. The horizontal dotted line is the isolated-two-bond prediction $t_{\ell,50}J=1.25$: the first three clocks vary weakly, whereas full many-body dressing strongly delays dominant-Schmidt-scale transport.}
\label{fig:s_bond_controls}
\end{figure}

The three tests in Fig.~\ref{fig:s_bond_controls} separate active-bond number in different initial states, global cut geometry, and internal many-body dressing. Conditioning random product states on the number of initially antiparallel cut bonds directly sorts the early response by active exchange bonds [Fig.~\ref{fig:s_bond_controls}(a)]. Matching two entanglement boundaries in an open central interval and a periodic half chain removes the apparent open--periodic difference [Fig.~\ref{fig:s_bond_controls}(b)]. Finally, keeping the cut couplings fixed while scaling every intra-half coupling by $g$ isolates the redistribution step [Fig.~\ref{fig:s_bond_controls}(c)]. The first three clocks vary weakly, whereas the dominant-Schmidt-scale transport time acquires a pronounced excess delay as the full many-body environment is restored.

\FloatBarrier
\subsection{Circuit and localization limits}

\begin{figure}[tbp]
\includegraphics[width=0.98\textwidth]{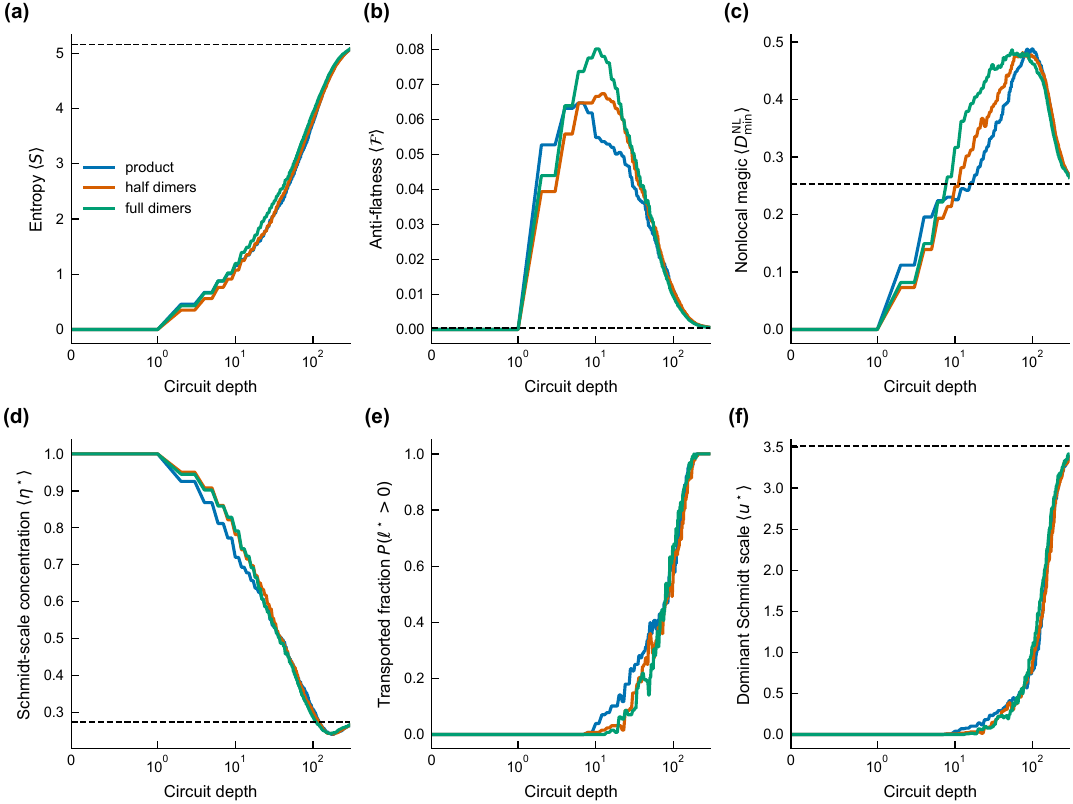}
\caption{Complete small-angle $U(1)$ circuit trajectories at $L=12$ and gate angle $\theta=0.25$, averaged over 128 circuit realizations for each product, half-dimer, and full-dimer initial reservoir defined in the text. (a) Entanglement entropy, (b) anti-flatness, (c) exact nonlocal magic, (d) Schmidt-scale concentration, (e) transported fraction, and (f) mean dominant Schmidt scale versus brick-wall circuit depth. The three reservoirs begin at distinct locations in spectrum space but approach the same late morphology. Black dashed lines show the fixed-charge Haar benchmarks for the observables in panels (a)--(d) and (f); panel (e) instead reports the probability of the binary event $\ell^\star>0$.}
\label{fig:s_circuit}
\end{figure}

We use a periodic $L=12$ qubit chain split into two six-site halves. Each half contains three neighboring one-particle dimers; a product reservoir leaves all six dimers in $|01\rangle$, a half-dimer reservoir prepares three randomly selected dimers in $(|01\rangle+|10\rangle)/\sqrt2$, and a full-dimer reservoir prepares all six in that Bell state. No dimer initially crosses the measured bipartition. In the full-dimer state, the two SWAP gates at the entanglement cuts transfer four Bell pairs across the bipartition, producing $2^4=16$ equal Schmidt weights: this is the flat rank-16 block. The transformation is $(S,\mathcal F,D_{\min}^{\rm NL},\eta^\star,u^\star):(0,0,0,1,0)\mapsto(4,0,0,1/2,3)$. Small-angle $U(1)$ circuits with $\theta=0.25$ instead generate roughness and magic barriers, then converge toward the fixed-charge Haar values. Figure~\ref{fig:s_circuit} shows that the three explicitly defined reservoirs approach the same late morphology by depth 300.

\begin{figure}[tbp]
\includegraphics[width=0.62\textwidth]{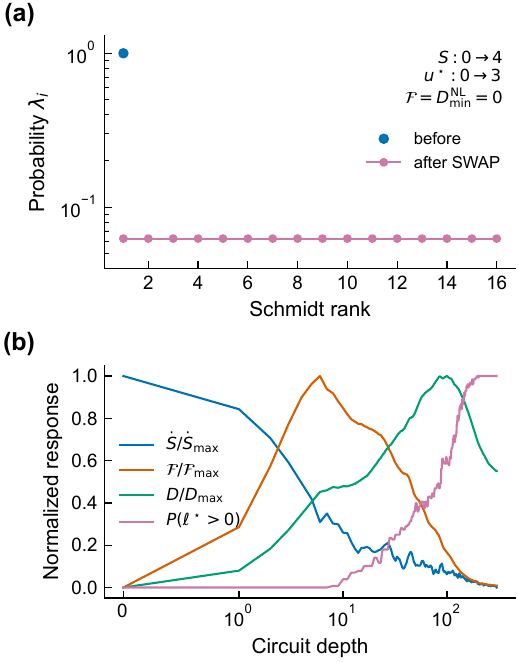}
\caption{Spectrum formation and transport controls for the periodic $L=12$ bipartition defined above. (a) Ordered Schmidt weights before and after the two cut-crossing SWAP gates transfer four Bell pairs across the bipartition. The initial rank-one spectrum becomes 16 equal weights, taking $(S,u^\star)$ from $(0,0)$ to $(4,3)$ while keeping $\mathcal F=D_{\min}^{\rm NL}=0$. (b) Normalized entropy-production rate, anti-flatness, and exact nonlocal magic, together with the transported fraction $P(\ell^\star>0)$, for the product-state $U(1)$ circuit at $\theta=0.25$, averaged over 128 realizations. The random gates successively reach maximal entropy production, roughness, magic, and dominant-Schmidt-scale transport; the SWAP protocol reaches a flat higher-rank block directly.}
\label{fig:s_build_move}
\end{figure}

Figure~\ref{fig:s_build_move} compares two operationally distinct circuit protocols. The SWAP protocol maps a rank-one spectrum to a flat rank-16 block, increasing entropy and the dominant Schmidt scale while keeping both anti-flatness and nonlocal magic exactly zero. Starting from a product state, the random circuit instead forms an uneven tail before its dominant Schmidt scale relocates.

\begin{figure}[tbp]
\includegraphics[width=0.98\textwidth]{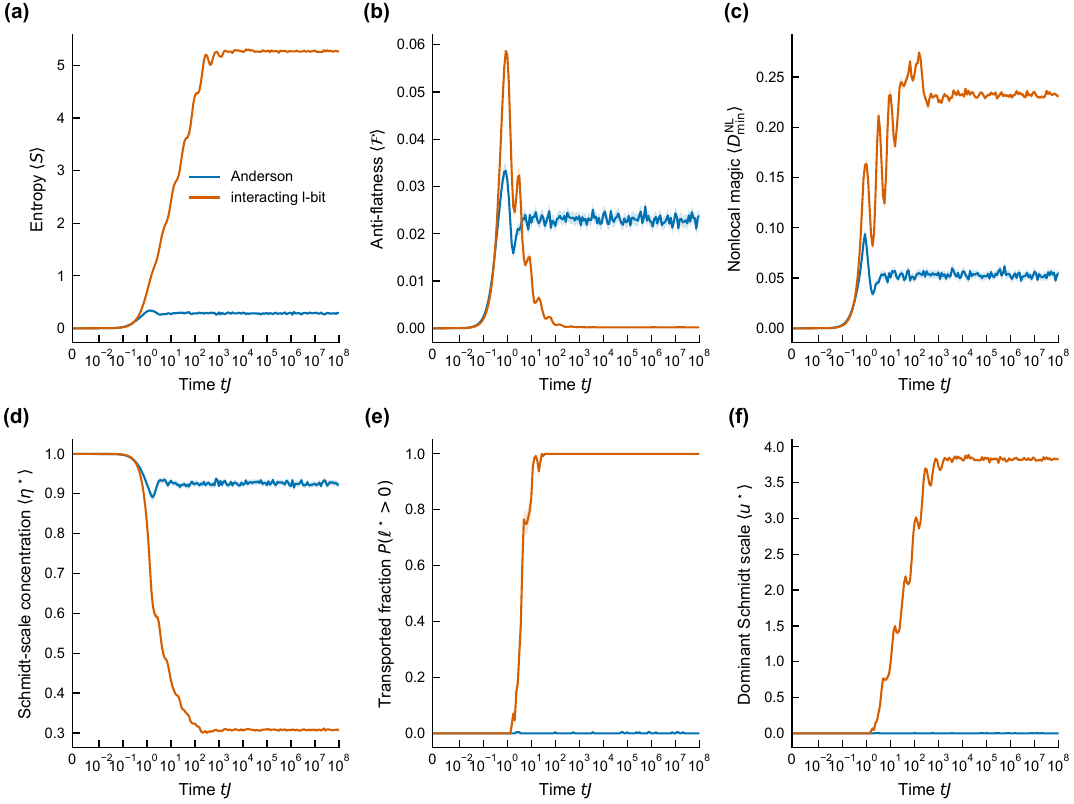}
\caption{Full comparison of noninteracting Anderson localization and interacting l-bit dephasing for $L=12$ through $tJ=10^8$. (a) Entanglement entropy, (b) anti-flatness, (c) exact nonlocal magic, (d) Schmidt-scale concentration, (e) transported fraction, and (f) mean dominant Schmidt scale. The Anderson control is an open-boundary random-field XX chain at $W=8$ evolved from the N\'eel state; the l-bit model has $J_0=\xi=1$, random-sign exponentially decaying interactions, and initial state $|+\rangle_\tau^{\otimes L}$. Curves are ensemble means and shading is the standard error over 256 Anderson and 128 l-bit realizations. The l-bit entropy grows approximately logarithmically over $1\lesssim tJ\lesssim10^2$ before finite-size saturation, while concentration loss and dominant-Schmidt-scale transport continue toward their finite-size endpoints; the Anderson spectrum remains head pinned.}
\label{fig:s_localization}
\end{figure}

Figure~\ref{fig:s_localization} displays the solvable localization limits analyzed in Sec.~6. The Anderson ensemble rapidly settles into a head-pinned area-law spectrum, whereas interacting l-bit dephasing from $|+\rangle_\tau^{\otimes L}$ produces approximately logarithmic entropy growth over $1\lesssim tJ\lesssim10^2$, followed by finite-size saturation, together with concentration loss, a transient roughness barrier, and dominant-Schmidt-scale transport.

\begin{figure}[tbp]
\includegraphics[width=0.98\textwidth]{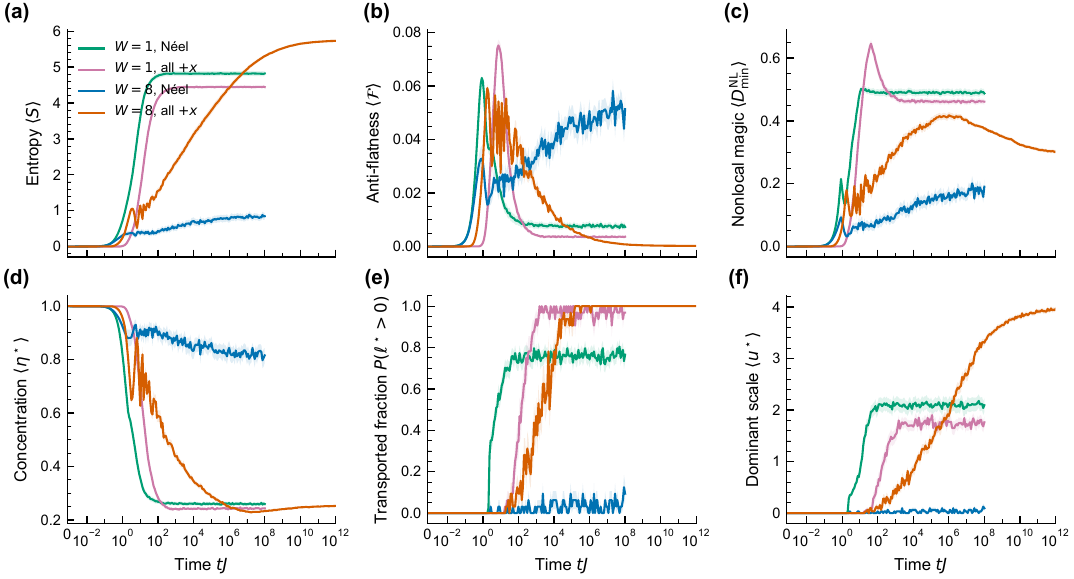}
\caption{Initial-state dependence of long-time Schmidt-spectrum dynamics at $L=14$. Green and pink curves show the $W=1$ N\'eel and physical-spin $|+x\rangle^{\otimes L}$ quenches through $tJ=10^8$, using 128 and 32 disorder realizations. Blue and orange curves show the corresponding $W=8$ quenches, using the same 32 disorder fields; the N\'eel trajectory ends at $tJ=10^8$ and the all-$+x$ trajectory continues to $tJ=10^{12}$. Panels show (a) entropy, (b) anti-flatness, (c) exact nonlocal magic, (d) Schmidt-scale concentration, (e) transported fraction, and (f) mean dominant Schmidt scale. Curves and bands are ensemble means and standard errors. The coherent all-$+x$ state activates extensive l-bit dephasing and dominant-Schmidt-scale transport even at strong disorder, whereas the strongly disordered N\'eel state remains close to the head-pinned l-bit configuration. }
\label{fig:s_initial_state}
\end{figure}

\textit{Initial-state dependence in the microscopic XXZ model.---}To test the role of the coherent l-bit initial condition directly, we repeat the $L=14$, $W=8$ XXZ evolution using the same 32 disorder realizations but replace the initial state by the physical-spin product state $|+x\rangle^{\otimes L}$. This state spans every conserved total-$S^z$ sector, so all 15 sectors are diagonalized separately and recombined with their relative phases intact. Figure~\ref{fig:s_initial_state} shows that this single protocol change qualitatively alters the spectrum dynamics. The transported fraction crosses $10\%$, $25\%$, $50\%$, and $90\%$ at the times $tJ=86.6,205.4,1333.5,$ and $20535.3$, respectively, and reaches one, while the matched $W=8$ dynamics with N\'eel initial state remains predominantly head pinned through $tJ=10^8$.

This contrast has a useful effective-l-bit interpretation, closely related to the long-time density-matrix construction used for symmetry restoration in MBL~\cite{liu2025mblmpemba}. Consider a uniformly tilted l-bit product state $\bigotimes_i[\cos(\theta/2)|\uparrow\rangle_{\tau,i}+\sin(\theta/2)|\downarrow\rangle_{\tau,i}]$. In the diagonal effective model, long-time dephasing suppresses off-diagonal configuration coherences under disorder averaging while preserving the local $\tau_i^z$ populations. Neglecting residual coherences and quasi-local dressing then gives the approximation
\begin{equation}
\overline{\rho_A(\infty)}\simeq\bigotimes_{i\in A}
\begin{pmatrix}
\cos^2(\theta/2)&0\\
0&\sin^2(\theta/2)
\end{pmatrix}.
\label{eq:s_lbit_tilted_plateau}
\end{equation}
At $\theta=0$, the right-hand side is a pure l-bit configuration and the diagonal Hamiltonian generates no dynamics. At $\theta=\pi/2$, corresponding to the maximally coherent $|+\rangle_\tau$ state, it reduces to $\mathbb I_A/2^{|A|}$. The effective picture therefore suggests why a physical all-$+x$ quench can acquire large entanglement and several thermal-like spectral features even deep in the localized regime, whereas a physical $z$-product state that lies close to one l-bit configuration remains head dominated. It does not predict an exactly flat Schmidt spectrum for the microscopic XXZ dynamics: the quasi-local rotation between physical spins and l-bits, finite-size residual coherences, and sample-to-sample fluctuations all generate visible deviations.

The endpoint structure must not be confused with thermal dynamics. Interactions at range $r$ dephase only at $t(r)\sim J_0^{-1}e^{r/\xi}$, so the active l-bit range grows as $r(t)\simeq\xi\ln(J_0t)$, as derived in Sec.~6. Entropy, concentration loss, and dominant-Schmidt-scale transport therefore develop over logarithmic time before finite-size saturation. In this coherent initial-state protocol, that logarithmically advancing dephasing front, rather than a small entanglement plateau, is the characteristic MBL signature.

Direct evolution of the all-$+x$ ensemble was continued to $tJ=10^{12}$. The final decade lies close to, but remains statistically distinguishable from, a separate random-eigenstate-phase estimate of the finite-size infinite-time typical plateau. Averaging 64 phase draws for each disorder realization gives
\begin{equation}
(S_\infty,\eta^\star_\infty,\lambda_{0,\infty},u^\star_\infty)
=(5.7463,0.25332,0.05672,3.9683),
\end{equation}
with standard errors $0.0306,0.00110,0.00179,$ and $0.0430$, respectively, and unit transported fraction. Randomizing eigenstate phases does not produce a Haar state because the energy-basis amplitudes and conserved-sector weights remain fixed. For comparison, the exact $L=14$ full-Haar Page entropy is $6.27870$, while finite-size Haar sampling gives $(\eta^\star,u^\star,D_{\min}^{\rm NL})=(0.28532,4.72910,0.24409)$. The all-$+x$ plateau therefore remains non-Haar despite its larger entropy than the $W=1$ N\'eel trajectory. Conversely, the weak-disorder N\'eel quench remains inside the half-filled sector and at fixed energy density.

\FloatBarrier
\subsection{Long-time XXZ dynamics and sample distributions}

\begin{figure}[tbp]
\includegraphics[width=0.98\textwidth]{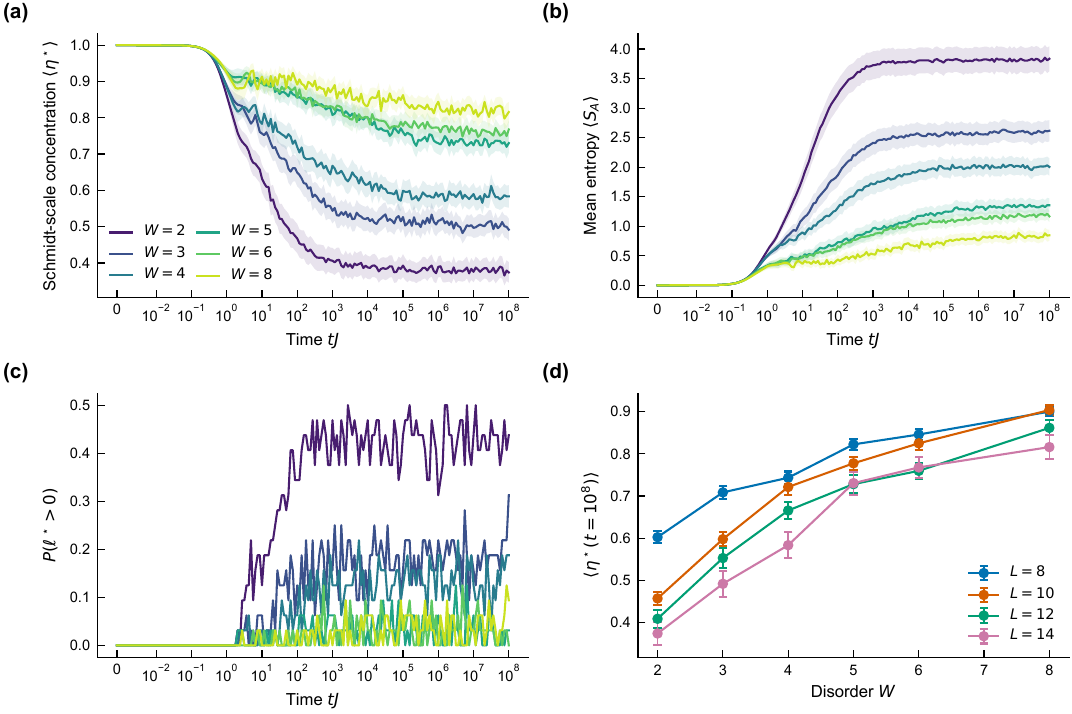}
\caption{Long-time random-field XXZ dynamics from N\'eel product states through $tJ=10^8$. (a)--(c) For $L=14$, the mean concentration $\langle\eta^\star\rangle$, entropy $\langle S\rangle$, and transported fraction $P(\ell^\star>0)$ are shown for $W=2,3,4,5,6,8$; shading denotes the standard error over 32 disorder realizations. (d) Endpoint concentration at $tJ=10^8$ versus disorder for $L=8,10,12,14$, with standard-error bars. Together, the time traces show continuing head erosion and the endpoint panel shows that increasing disorder and size suppress the resulting dominant-Schmidt-scale transport. The four sizes contain 128, 96, 64, and 32 realizations per disorder value.}
\label{fig:s_long_time}
\end{figure}

The XXZ data connect the solvable localization controls to the interacting spin chain. Figure~\ref{fig:s_long_time} extends direct fixed-sector evolution to $tJ=10^8$ for additional disorder strengths. Strong disorder continues to gain entropy and lose head concentration over many decades, while a large fraction of samples remain pinned at $\ell^\star=0$. Near-saturation of entropy only says that the total spectral spread changes slowly; the largest Schmidt weights can still redistribute on longer dephasing scales, to which the head-weighted anti-flatness remains sensitive.

\FloatBarrier

\begin{figure}[tbp]
\includegraphics[width=0.98\textwidth]{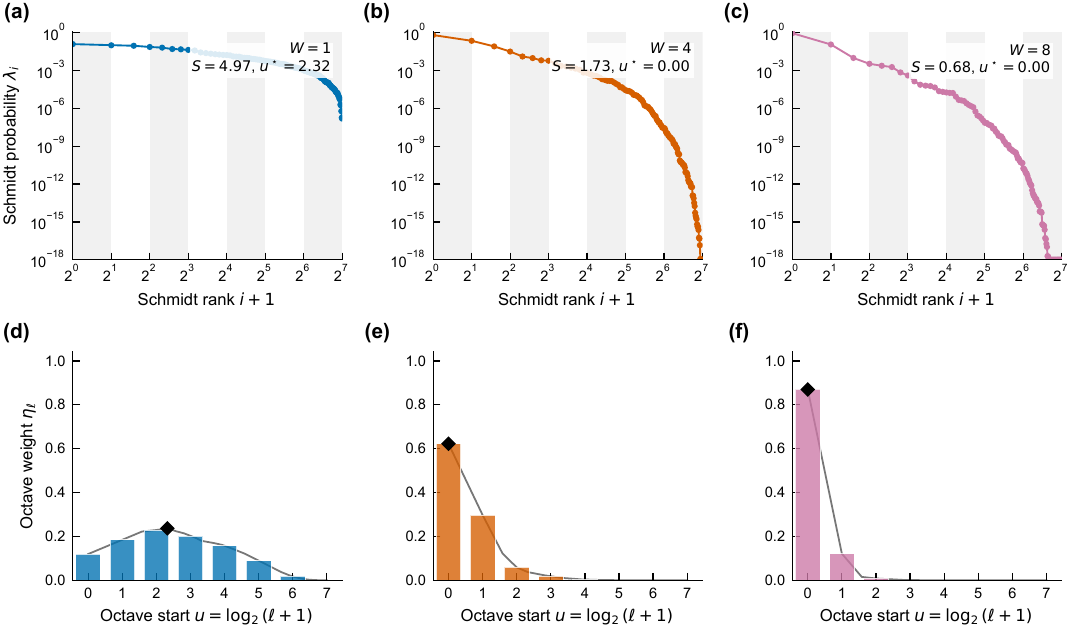}
\caption{Representative individual Schmidt spectra and their logarithmic-rank coarse graining for $L=14$ N\'eel quenches at $tJ=10^8$; they are representative examples rather than ensemble averages. (a)--(c) Ordered Schmidt probabilities $\lambda_i$ versus the one-based rank $i+1$, with both axes logarithmic. Alternating gray bands deliberately select the power-of-two-aligned intervals $2^k\leq i+1<2^{k+1}$, a nonoverlapping subset of the sliding windows used in the analysis. (d)--(f) The bars sum the probability in this illustrative subset, so the bar at integer $u=k$ equals $\eta_{2^k-1}=\sum_{i=2^k-1}^{\min(2^{k+1}-2,r-1)}\lambda_i$. The gray line is the complete overlapping profile $\eta_\ell$, evaluated for every integer starting rank at $u=\log_2(\ell+1)$, and the black diamond marks its maximum $(u^\star,\eta^\star)$. The broad $W=1$ spectrum places its dominant window at a higher rank scale, while the $W=4$ and $8$ examples remain head dominated.}
\label{fig:s_representative_spectra}
\end{figure}

The coarse graining from an ordered Schmidt spectrum to logarithmic-rank weight is shown directly in Fig.~\ref{fig:s_representative_spectra}. For visual clarity, its alternating bands and bars select the power-of-two-aligned windows $I_{2^k}$, which form a nonoverlapping cover; the optimization of $\eta^\star$ still uses every overlapping window $I_q$. At the same size and time, weak disorder produces a broad spectrum whose largest octave begins away from the leading Schmidt weight, whereas the crossover and strongly localized examples retain most probability in the head octave.

\FloatBarrier

\begin{figure}[tbp]
\includegraphics[width=0.98\textwidth]{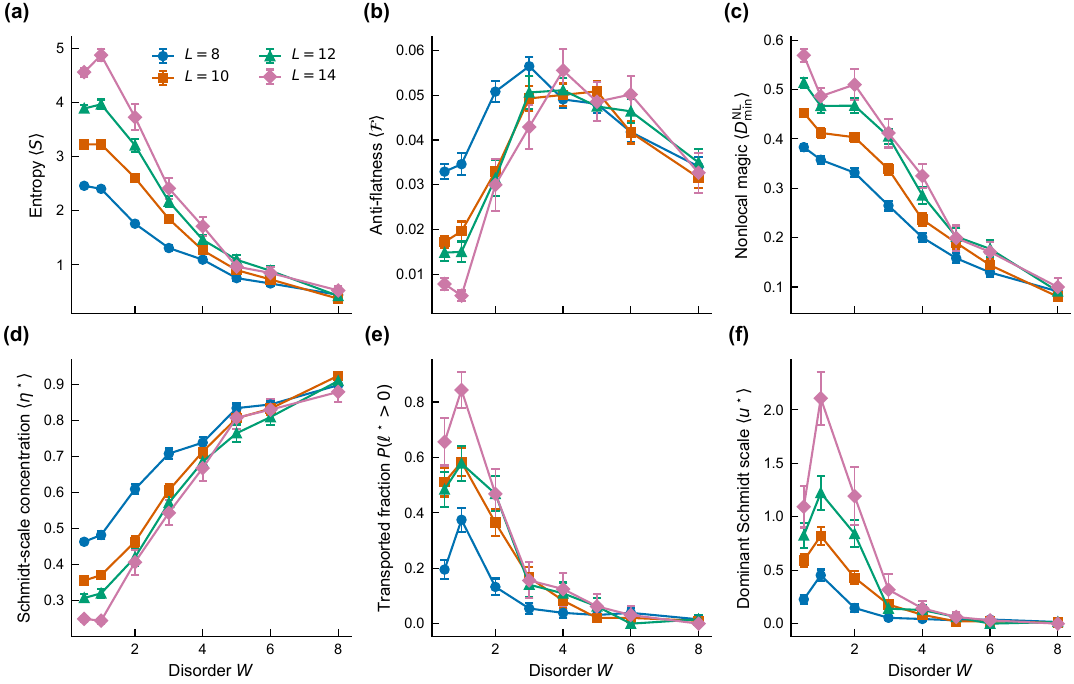}
\caption{Disorder and size dependence of the six spectral observables at the common observation time $tJ=10^3$. Panels show (a) entropy, (b) anti-flatness, (c) exact nonlocal magic, (d) Schmidt-scale concentration, (e) transported fraction, and (f) mean dominant Schmidt scale for $L=8,10,12,14$. Error bars are standard errors over 128, 96, 64, and 32 disorder realizations, respectively. The intermediate-disorder dome in anti-flatness contrasts with the overall suppression of entropy and nonlocal magic and the increase of Schmidt-scale concentration toward large $W$; dominant-Schmidt-scale transport is rapidly suppressed through the crossover.}
\label{fig:s_disorder_size}
\end{figure}

To compare every observable on one matched grid, Fig.~\ref{fig:s_disorder_size} takes a fixed-time cut at $tJ=10^3$. Beyond the weak-disorder plateau, entropy decreases with $W$ and grows strongly with $L$ on the thermal side, while this size growth is largely lost at strong disorder. Concentration shows the opposite trend, increasing toward one with $W$ and decreasing with $L$ primarily for $W\lesssim4$. The transported fraction and mean location are largest at weak disorder and collapse through the $W=3$--$4$ crossover. Anti-flatness is nonmonotonic, with a broad maximum at intermediate disorder because both the broadly thermal spectrum and the nearly rank-one localized spectrum are relatively flat. Exact nonlocal magic grows with size on the thermal and crossover sides and is strongly suppressed toward large $W$, with small nonmonotonic finite-time variations at weak disorder. These trends distinguish the total amount of entanglement, head--tail roughness, dyadic nonflatness, concentration, and dominant-Schmidt-scale transport. Because localized trajectories still drift beyond $tJ=10^3$, this matched comparison is a dynamical cross-section rather than a stationary phase diagram; Fig.~\ref{fig:s_long_time} separately displays the available $tJ=10^8$ endpoints.

\begin{figure}[tbp]
\includegraphics[width=0.98\textwidth]{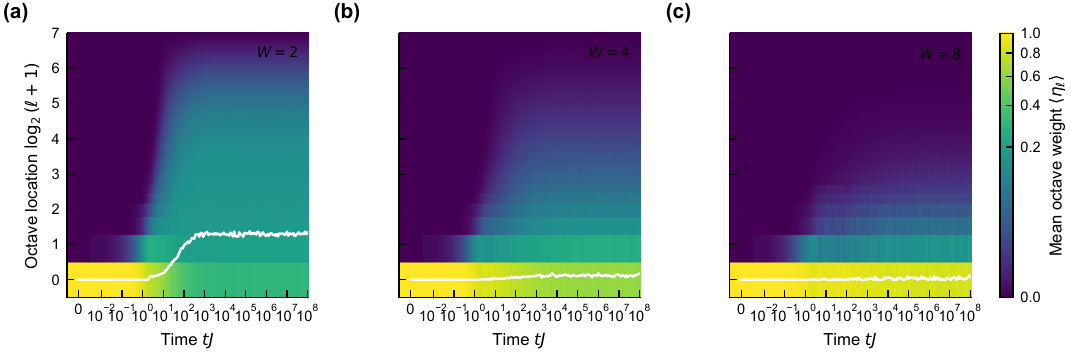}
\caption{Disorder-resolved evolution of the complete sliding-octave profile for $L=14$ N\'eel quenches. Panels (a)--(c) use $W=2$, $4$, and $8$ dynamics through $tJ=10^8$, averaged over 128, 128, and 32 disorder realizations, respectively. The horizontal coordinate is time, the vertical coordinate is the sliding window start $u=\log_2(\ell+1)$, and color gives the ensemble-mean window mass $\langle\eta_\ell\rangle$. The white curve is the mean dominant scale $\langle u^\star\rangle$. It rises clearly for $W=2$, moves only weakly for $W=4$, and remains near the head for $W=8$.}
\label{fig:s_heatmaps}
\end{figure}

\begin{figure}[tbp]
\includegraphics[width=0.98\textwidth]{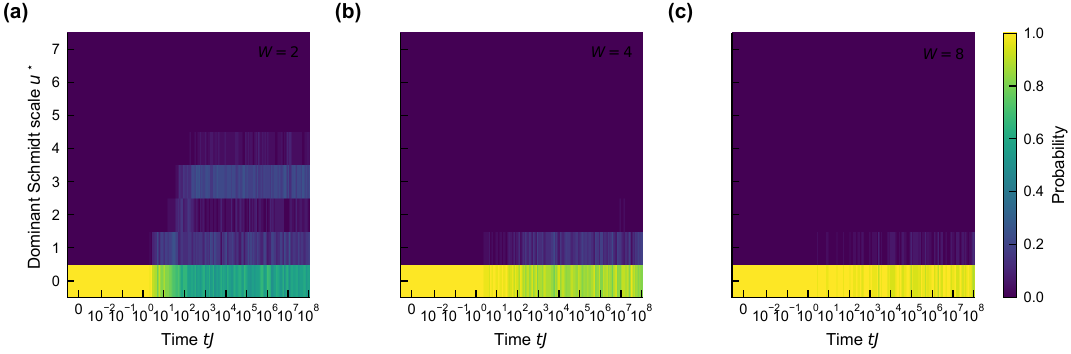}
\caption{Full distribution of the dominant Schmidt scale $u^\star=\log_2(\ell^\star+1)$ for $L=14$ N\'eel quenches. Panels (a)--(c) show $W=2$, $4$, and $8$, respectively, from 32 disorder realizations through $tJ=10^8$. For visualization, the values of $u^\star$ are grouped into unit-width bins; color is the empirical probability in each bin at each time. The persistent bright bin containing $u^\star=0$ exposes the zero-inflated, non-Gaussian distribution, while weight in higher-$u^\star$ bins records how far the moved realizations travel. }
\label{fig:s_location}
\end{figure}

The full octave heatmaps in Fig.~\ref{fig:s_heatmaps} distinguish gradual head erosion from migration of the largest window. Figure~\ref{fig:s_location} further shows that $\ell^\star$ is zero-inflated and strongly non-Gaussian. We therefore report the transported fraction $P(\ell^\star>0)$. At $L=14$ and $tJ=10^8$, the transported fractions for $W=2,4,8$ are $0.50\pm0.04$, $0.16\pm0.03$, and $0.09\pm0.05$, respectively, with binomial standard errors from $N=128,128,$ and $32$ realizations.

\begin{figure}[tbp]
\includegraphics[width=0.98\textwidth]{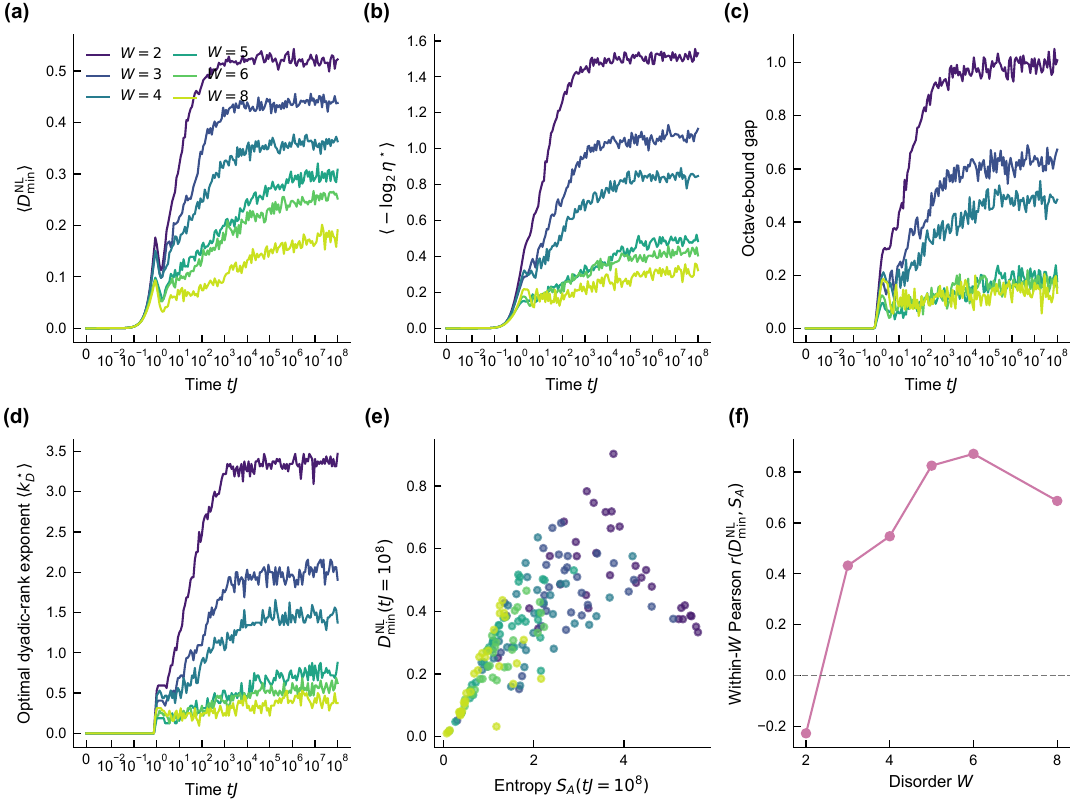}
\caption{Exact nonlocal magic and its relation to concentration and entropy in the long-time XXZ data. Panels (a)--(d) show, for $L=14$ and $W=2,3,4,5,6,8$, the disorder means of (a) $D_{\min}^{\rm NL}$, (b) its octave upper bound $-\log_2\eta^\star$, (c) the gap between that bound and the exact value, and (d) the optimizing dyadic-rank exponent $k_D^\star$ versus time. Here $2^{k_D^\star}$ is the rank of the flat Schmidt spectrum that maximizes the fidelity in Eq.~\eqref{eq:s_magic}; changes in $k_D^\star$ identify switches of the closest dyadic-flat rank sector. Each curve averages 32 realizations. (e) Sample-resolved endpoint scatter of $D_{\min}^{\rm NL}$ against $S_A$ at $tJ=10^8$, with color indicating disorder. (f) Pearson correlation $r(D_{\min}^{\rm NL},S_A)$ computed separately within each disorder ensemble; the dashed line marks zero.}
\label{fig:s_magic}
\end{figure}

Figure~\ref{fig:s_magic} compares exact nonlocal magic with its octave bounds. Across all $L=14$, $tJ=10^8$ endpoints from the six disorder ensembles $W=2$--8, the correlations are $r(\eta^\star,S)=-0.9417$, $r(D_{\min}^{\rm NL},S)=0.6712$, and $r(D_{\min}^{\rm NL},\eta^\star)=-0.6481$. These combined values mainly reflect how the ensemble centers move with disorder. Within a fixed-$W$ ensemble the magic--entropy correlation can have either sign because the closest dyadic-flat rank sector can change along the spectral trajectory [Eq.~\eqref{eq:s_headtail_magic}].

\FloatBarrier
\section{Symmetry-resolved random-state benchmarks}

The late-time thermal reference must respect the conserved total magnetization of the XXZ quench. This section derives the corresponding concentration and location baselines and quantifies the shift from an unconstrained Haar state.

\begin{figure}[tbp]
\includegraphics[width=0.98\textwidth]{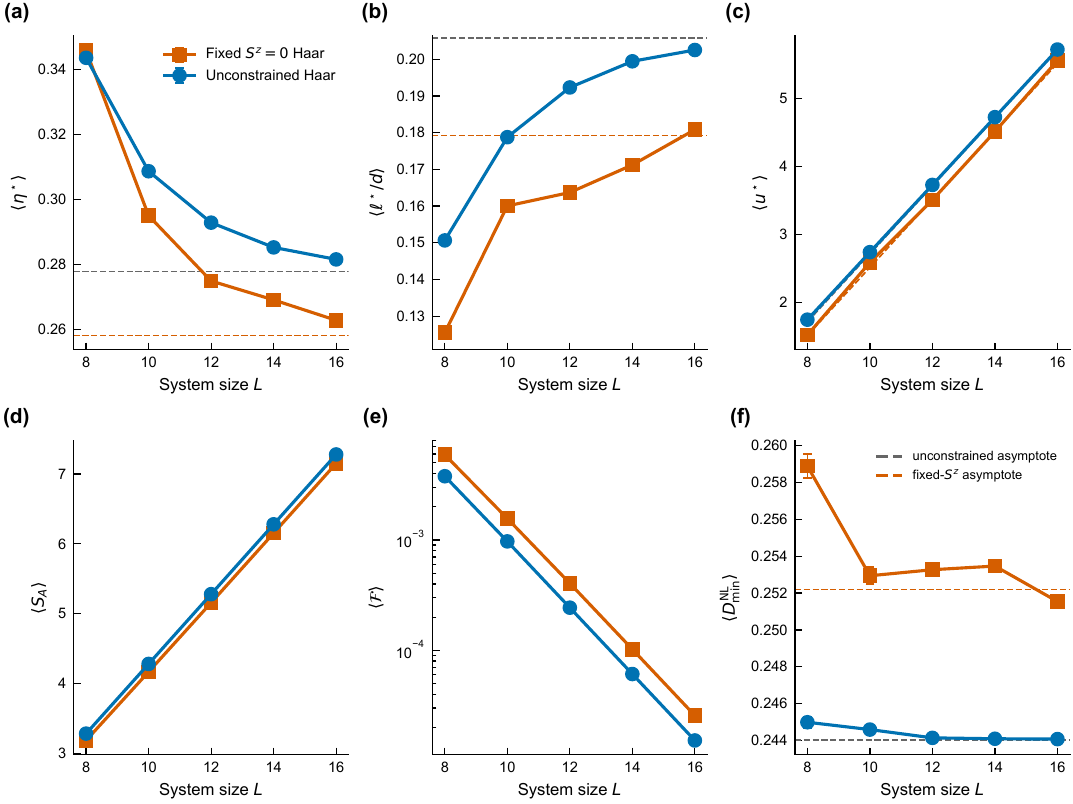}
\caption{Unconstrained and half-filled Haar benchmarks from exact complex block-Wishart sampling. Orange squares denote Haar states in the fixed $S^z=0$ sector and blue circles denote unconstrained bipartite Haar states for $L=8,10,12,14,16$; error bars are standard errors from 4096, 2048, 1024, 512, and 256 samples, respectively. Panels show (a) Schmidt-scale concentration $\langle\eta^\star\rangle$, (b) normalized maximizing index $\langle\ell^\star/d\rangle$ with $d=2^{L/2}$, (c) dominant Schmidt scale $\langle u^\star\rangle$, (d) von Neumann entropy, (e) anti-flatness on a logarithmic scale, and (f) exact nonlocal magic. Gray dashed references in panels (a)--(c) and (f) are the unconstrained-Haar limits in Eqs.~\eqref{eq:s_full_haar} and \eqref{eq:s_full_haar_magic}. Orange dashed references in the same panels are the fixed-charge limits from Eq.~\eqref{eq:s_u1_haar}.}
\label{fig:s_haar}
\end{figure}

For an unconstrained Haar-random state on $\mathbb C^d\otimes\mathbb C^d$, where $d=\dim\mathcal H_A=\dim\mathcal H_B$ is the Hilbert-space dimension of either half, the rescaled eigenvalues $x=d\lambda$ follow the Marchenko--Pastur (MP) density~\cite{marcenko1967distribution,lubkin1978entropy,page1993average,zyczkowski2001induced,sommers2004statistical,nadal2010phase,nadal2011statistical}
\begin{equation}
\rho_{\rm MP}(x)=\frac{1}{2\pi}\sqrt{\frac{4-x}{x}},\qquad 0<x<4.
\end{equation}
Let $Q(x)=\int_x^4\rho_{\rm MP}(y)\,dy$ be the fraction of eigenvalues above threshold $x$, and let $W(x)=\int_x^4y\rho_{\rm MP}(y)\,dy$ be their total Schmidt weight. The substitution $x=4\cos^2\theta$ gives
\begin{align}
Q(\theta)&=\frac{2\theta-\sin2\theta}{\pi},\\
W(\theta)&=\frac{2\theta-\tfrac12\sin4\theta}{\pi}.
\end{align}
An octave beginning at the rank fraction $\alpha=\ell/d$ contains ranks from $\alpha d$ to $2\alpha d$ and hence has asymptotic mass $\eta(\alpha)=W[\theta(2\alpha)]-W[\theta(\alpha)]$, where $Q[\theta(\alpha)]=\alpha$. Differentiating with respect to $\alpha$ uses $dW/dQ=x$ and yields the stationarity condition
\begin{equation}
x(\alpha)=2x(2\alpha).
\end{equation}
Solving this condition together with $Q[\theta(\alpha)]=\alpha$ and $Q[\theta(2\alpha)]=2\alpha$ gives $\alpha^\star=0.2058302$ and $\eta(\alpha^\star)=0.2779017$. Since $\ell^\star=\alpha^\star d+o(d)$, $u^\star=\log_2d+\log_2\alpha^\star+o(1)$, yielding
\begin{equation}
\frac{\ell^\star}{d}\longrightarrow0.2058302,\qquad \eta^\star\longrightarrow0.2779017,\qquad u^\star=\log_2d-2.280473+o(1).
\label{eq:s_full_haar}
\end{equation}
The same MP law gives the asymptotic exact nonlocal magic. Define the upper-tail Schmidt-amplitude integral
\begin{equation}
A(x)=\int_x^4\sqrt y\,\rho_{\rm MP}(y)\,dy.
\label{eq:s_full_haar_amplitude}
\end{equation}
With $x=4\cos^2\theta$, the rank and amplitude tails are
\begin{equation}
Q(\theta)=\frac{2\theta-\sin2\theta}{\pi},\qquad
A(\theta)=\frac{8\sin^3\theta}{3\pi}.
\label{eq:s_full_haar_tails}
\end{equation}
A dyadic-flat comparison state of rank $m=\beta d$, with $\beta=1,1/2,1/4,\ldots$, retains the largest fraction $\beta$ of the ordered spectrum. If $Q(\theta_\beta)=\beta$, self-averaging of the empirical spectrum gives $\sum_{j<m}\sqrt{\lambda_j}=\sqrt d\,A(\theta_\beta)+o(\sqrt d)$ and hence
\begin{equation}
F_\beta=\frac{A(\theta_\beta)^2}{\beta}
=\frac{64\sin^6\theta_\beta}{9\pi^2\beta},\qquad
2\theta_\beta-\sin2\theta_\beta=\pi\beta.
\label{eq:s_full_haar_fidelity}
\end{equation}
The continuous function $F[Q(\theta)]$ is unimodal. Its derivative has the sign of $g(\theta)=3\theta\cos\theta-\sin\theta(1+2\cos^2\theta)$, for which $g'(\theta)=3\sin\theta(\sin2\theta-\theta)$. Thus $g$ first increases and then decreases, crossing zero once at $\theta=1.2729796061$, corresponding to $\beta=0.6318215649$. Consequently $F(\beta)$ increases throughout $0<\beta\leq1/2$. The best dyadic candidate on that interval is therefore $\beta=1/2$, and it also exceeds the only larger candidate because $F_{1/2}=0.8443880802\ldots>F_1=64/(9\pi^2)=0.7205061948\ldots$. The dyadic optimum is $\beta^\star=1/2$. Writing
\begin{equation}
2\theta_\star-\sin2\theta_\star=\frac{\pi}{2},\qquad
\theta_\star=1.1549407300\ldots,
\end{equation}
we obtain
\begin{equation}
F_{\rm NL}\longrightarrow\frac{128\sin^6\theta_\star}{9\pi^2}
=0.8443880802\ldots,\qquad
D_{\min}^{\rm NL}\longrightarrow0.2440218818\ldots\ \mathrm{bits}.
\label{eq:s_full_haar_magic}
\end{equation}
These are self-averaging $d\to\infty$ limits for balanced unconstrained Haar states, rather than exact finite-$d$ ensemble averages. The $L=16$ numerical value $D_{\min}^{\rm NL}=0.2440718$ bits is already within $5.0\times10^{-5}$ bits of Eq.~\eqref{eq:s_full_haar_magic}. We next impose fixed total charge; this changes the limiting spectral density and shifts the magic limit to $0.2521994$ bits.

At half filling, let the full chain contain $2n$ spins, the subsystem contain $n$ spins, and the total number of up spins be $n$. The reduced state decomposes as $\rho_A=\bigoplus_q\rho_A^{(q)}$. The block with $q$ up spins in $A$ is a square Wishart matrix of dimension $d_q=\binom nq$, and its mean trace is the hypergeometric weight
\begin{equation}
p_q=\frac{d_q^2}{\binom{2n}{n}}.
\end{equation}
Thus the ordered spectrum is a scale mixture of MP blocks rather than one MP law. A central-limit expansion about $q=n/2$ introduces the continuous sector coordinate $z=(q-n/2)/\sqrt n$. The asymptotic density of Schmidt values per unit $z$ is $w(z)=\sqrt{2/\pi}\,e^{-2z^2}$, their relative eigenvalue scale is $s(z)=\sqrt2\,e^{-2z^2}$, and their Schmidt-weight density is $w(z)s(z)$. Define the MP tail integrals
\begin{equation}
Q_0(x)=\int_x^4\rho_{\rm MP}(u)\,du,\quad
Q_1(x)=\int_x^4u\rho_{\rm MP}(u)\,du,\quad
Q_{1/2}(x)=\int_x^4\sqrt{u}\rho_{\rm MP}(u)\,du.
\end{equation}
The rank fraction, probability weight, and square-root amplitude above a global threshold $y$ are then
\begin{equation}
\mathcal Q_a(y)=2\int_0^\infty dz\,w(z)s(z)^aQ_a\!\left(\frac{y}{s(z)}\right),
\qquad a=0,1,\frac12.
\end{equation}
The normalizations $\mathcal Q_0(0)=\mathcal Q_1(0)=1$ check the count and probability measures. The threshold $y(\alpha)$ is defined by $\mathcal Q_0[y(\alpha)]=\alpha$. Exactly as above, the octave mass is $\mathcal Q_1[y(2\alpha)]-\mathcal Q_1[y(\alpha)]$, and its stationary point satisfies $y(\alpha)=2y(2\alpha)$. Numerical quadrature of these one-dimensional integrals gives $\alpha^\star=0.1792176$ and $\eta^\star=0.2582366$.

For nonlocal magic, a dyadic prefix occupying rank fraction $\beta=2^{-k}$ has asymptotic fidelity
\begin{equation}
F_\beta=\frac{\mathcal Q_{1/2}[y(\beta)]^2}{\beta}.
\end{equation}
Maximizing over the dyadic sequence $\beta=1,1/2,1/4,\ldots$ gives $D_{\min}^{\rm NL}=-\log_2\max_\beta F_\beta=0.2521994$ bits. Collecting the symmetry-resolved constants yields
\begin{equation}
\frac{\ell^\star}{d}\longrightarrow0.1792176,\qquad \eta^\star\longrightarrow0.2582366,\qquad D_{\min}^{\rm NL}\longrightarrow0.2521994\ \text{bits}.
\label{eq:s_u1_haar}
\end{equation}
At $L=16$, the fixed-charge values $(\ell^\star/d,\eta^\star,D_{\min}^{\rm NL})=(0.180832,0.262912,0.251540)$ approach the asymptotic prediction in Eq.~\eqref{eq:s_u1_haar}.

The symmetry-resolved curve in Fig.~\ref{fig:s_haar} is the appropriate infinite-temperature endpoint for the charge-conserving random circuits. The N\'eel quench instead has finite energy density, so even if it thermalizes its endpoint is canonical rather than Haar; the fixed-charge Haar values therefore provide a random-state reference, not a predicted endpoint, for weak-disorder XXZ dynamics. The order-one $\eta^\star$ of the random-state benchmark also separates thermal random states from universal embezzling families, whose defining spectral condition is $\eta^\star\to0$~\cite{vanDam2003embezzling,cleve2017perfect,zanoni2024complete,vanLuijk2025critical}.

\bibliography{ref}